\documentclass[12pt]{article}
\usepackage{epsfig}
\usepackage{amsmath}
\usepackage{hhline}
\usepackage{amssymb}
\usepackage{times}
\usepackage{cite}
\usepackage{authblk}
\usepackage{xcolor}

\usepackage{url}

\usepackage[figuresright]{rotating}
\usepackage{hyperref} 

\usepackage{caption}
\usepackage{float} 
\begin{document}

\begin{titlepage}

\vspace*{2cm}

\begin{center}
\begin{Large}

{\boldmath \bf
    Production of protons, deuterons and tritons in argon-nucleus
    interactions at 3.2~A~GeV
}

\vspace{0.5cm}

BM$@$N Collaboration

\end{Large}
\end{center}

\vspace{0.5cm}

{\noindent
S.\,Afanasiev$^1$, G.\,Agakishiev$^1$, A.\,Aleksandrov$^9$, E.\,Aleksandrov$^1$, I.\,Aleksandrov$^1$, P.\,Alekseev$^{1,3}$,  K.\,Alishina$^1$, V.\,Astakhov$^1$, T.\,Aushev$^5$, V.\,Azorskiy$^1$
% ...
}


{\small \noindent

\noindent
$1${\ Joint Institute for Nuclear Research (JINR), Dubna, Russia}

\noindent
$2${\ Institute for Nuclear Research of the RAS (INR RAS), Moscow, Russia}

\noindent
$3${\ Kurchatov Institute, NRC, Moscow, Russia}

\noindent
$4${\ Lebedev Physical Institute of the Russian Academy of Sciences (LPI RAS), Moscow, Russia}

\noindent
$5${\ Moscow Institute of Physics and Technology (MIPT), Moscow, Russia}

\noindent
$6${\ National Research Nuclear University MEPhI, Moscow, Russia}

\noindent
$7${\ National Research University Higher School of Economics (HSE University), Moscow, Russia}

\noindent
$8${\ Skobeltsyn Institute of Nuclear Physics, Moscow State University (SINP MSU), Moscow, Russia}

\noindent
$9${\ St Petersburg University (SPbU), St Petersburg, Russia}

\noindent
$10${\ Plovdiv University ``Paisii Hilendarski'', Plovdiv, Bulgaria}

\noindent
$11${\ Key Laboratory of Nuclear Physics and Ion-Beam Application (MOE), Institute of Modern Physics, Fudan University, Shanghai, China}

\noindent
$12${\ Institute of Physics and Technology, Satbayev University, Almaty, Kazakhstan}

\noindent
$13${\ Physical-Technical Institute of Uzbekistan Academy of Sciences (PhTI of UzAS), Tashkent, Uzbekistan}

\noindent
$14${\ Institute of Mechanics at the Bulgarian Academy of Sciences (IMech-BAS), Sofia, Bulgaria}

\noindent
$15${\ Weizmann Institute of Science, Rehovot, Israel}
}

\newpage

\begin{abstract}
\noindent 
{Results of the BM$@$N experiment at the Nuclotron/NICA complex on the production of protons, deuterons 
and tritons in interactions of an argon beam of 3.2 A GeV with fixed targets of C, Al, Cu, Sn and Pb 
are presented. Transverse mass spectra, rapidity distributions and 
multiplicities of protons, deuterons and tritons  are measured. The results
are treated within a coalescence approach and compared with predictions of theoretical 
models and with other measurements.}
\end{abstract}

\vspace{1cm}

\end{titlepage}

\section{Introduction}
\label{sect1}

BM$@$N (Baryonic Matter at Nuclotron) is the first operational experiment at
the Nuclotron/NICA accelerator complex. The Nuclotron provides beams
of a variety of particles, from protons up to gold ions, with kinetic energy
in the range from 1 to 6~A~GeV for light ions with Z/A ratio
of $\sim0.5$ and up to 4.5~A~GeV for heavy ions with Z/A ratio
of $\sim0.4$. At these energies, the nucleon density in the fireball created 
in collisions of heavy ions with fixed targets is 3--4 times
higher than the nuclear saturation density~\cite{Friman}, thus allowing one to
study heavy-ion interactions in the high-density baryonic matter 
regime~\cite{Cleymans,Fuchs,NICAWhitePaper,BMN_CDR}.

During the commissioning phase, BM$@$N, in a configuration with limited 
phase-space coverage, collected its first data with beams of carbon, argon and krypton
ions \cite{BMN_QM, BMN_SQM}. In the first physics publication, BM$@$N reported 
studies of $\pi^+$ and $K^+$ production in argon-nucleus interactions~\cite{BMN_piKpaper}.
 This paper presents results
on  proton, deuteron and triton production in 3.2~A~GeV argon-nucleus interactions.

At the Nuclotron energies, baryon transfer over finite rapidity distances (baryon
stopping~\cite{stop_1}) plays an important role~\cite{stop_2}\---\cite{stop_4}. The baryon 
density achieved in high-energy nuclear collisions is a crucial quantity that governs the 
reaction dynamics and the overall system evolution, including eventual phase transitions. 
The baryon rapidity distributions in heavy ion collisions for different combinations of 
projectile and target as well as at different impact parameters provide essential constraints
on the dynamical scenarios of baryon stopping.  The BM$@$N experimental setup allows 
for the measurement of the distribution of protons and light nuclei ($d, t$) over 
the rapidity interval [1.0--2.2]. This rapidity range is wide enough to include not 
only the midrapidity (rapidity of the nucleon-nucleon center-of-mass (CM) system 
is $y_{CM}$ = 1.08) but also the beam rapidity region ($y_{beam}$\,=\,2.16), in contrast to 
the collider experiments focused mainly on in the mid-rapidity
region. Another advantage of BM$@$N consists in the coverage of a wide interval 
of transverse momenta ($p_T$) of produced nuclear clusters (light nuclei). 
This makes possible to determine the general shape of the rapidity density 
distribution and derive information about the rapidity shift and energy loss 
in nucleus-nucleus collisions.

Nuclear cluster production allows one  to estimate the nucleon phase-space density attained 
in the reaction~\cite{na44_phasespace}. It governs the overall evolution of the reaction 
process and may provide information about freeze-out conditions and entropy production 
in relativistic nucleus-nucleus interactions. The nucleon phase space density can be obtained from the 
ratio of deuteron and proton abundances. One of the goals of this work is to study the 
particle phase-space density evolution in Ar+A collisions for different projectile-target 
combinations and as a function of collision centrality.   

In collisions of heavy nuclei at relativistic energies, a significant fraction of the 
initial kinetic energy transforms into particle production and thermal excitation of
matter. Various dynamical models, including those based on hydrodynamics, have demonstrated 
that the entropy per baryon $S/A$ created during the initial interaction stage remains 
constant during the subsequent evolution of the system~\cite{entr1, entr2}. Therefore, 
entropy production data provide insight not only into the nucleon phase-space density 
at the final stage of the reaction (freeze-out), but also into the properties of the 
medium during the hot and dense stage. It is also the aim of this work to investigate 
the entropy evolution in the reaction zone with system size in argon-nucleus collisions 
and compare BM$@$N results with results of other experiments.

The few MeV binding energies of the deuteron and the triton are much lower than the 
freeze-out temperatures estimated to be above 100 MeV. These
light clusters are therefore not expected to survive through
the high density stages of the collision. The deuterons and
tritons observed in the experiment are  emitted
at the end of the freeze-out process, carrying information about
this late stage of the collision.

Light cluster production in low-energy heavy-ion collisions is well described by 
a simple coalescence model~\cite{Coal1,Coal2,Coal3,Sat81}
based on the distributions of their constituents (protons and
neutrons) and a coalescence parameter $B_A$ related to the cluster mass number A.
In order to describe heavy-ion collisions at high energies, the simple coalescence model 
has been modified to account for the nucleon phase space
distributions at the freeze-out and also for the strength of the momentum-space
correlations induced by collective flow~\cite{Scheibl}.
 In central heavy-ion collisions, the pressure gradient in
the system generates strong transverse radial flow. Therefore, nucleon clusters 
inside a collective velocity field acquire additional
momenta proportional to the masses of these clusters. 

The paper is organized as follows: the experimental setup is described in 
Section~\ref{sect2}, the event reconstruction is detailed in Section~\ref{sect3}, 
and the evaluation of proton, deuteron, and triton reconstruction efficiency 
is presented in Section~\ref{sect4}. The methodology to define collision centrality 
classes is explained in Section~\ref{sect44}. The evaluation of the cross sections, 
multiplicities, and systematic uncertainties is addressed in Section~\ref{sect5}. 
The transverse mass and rapidity distributions of protons, deuterons, and tritons 
are presented in Section~\ref{sect6}. The BM$@$N results are compared with 
predictions from the DCM-SMM~\cite{DCM_QGSM,DCM_SMM} and PHQMD~\cite{PHQMD} models. 
The ratios of the transverse momentum distributions of deuterons and tritons to 
protons are treated within a coalescence approach in Section~\ref{sect66}. 
The results are compared with other experimental data on nucleus-nucleus interactions.
 Results on baryon rapidity loss in argon-nucleus interactions are presented in 
 Section~\ref{section_stopping}. The compound ratios of yields of protons and 
 tritons to deuterons are presented in Section~\ref{sect67}. Finally, a summary 
 is given in Section~\ref{sect7}.

\section{Experimental setup}
\label{sect2}

\begin{figure}[htb]
\begin{center}
\vspace{1.5cm}
\hspace{-3.8cm} \includegraphics[width=0.028\textwidth,bb=550 0 590 120]{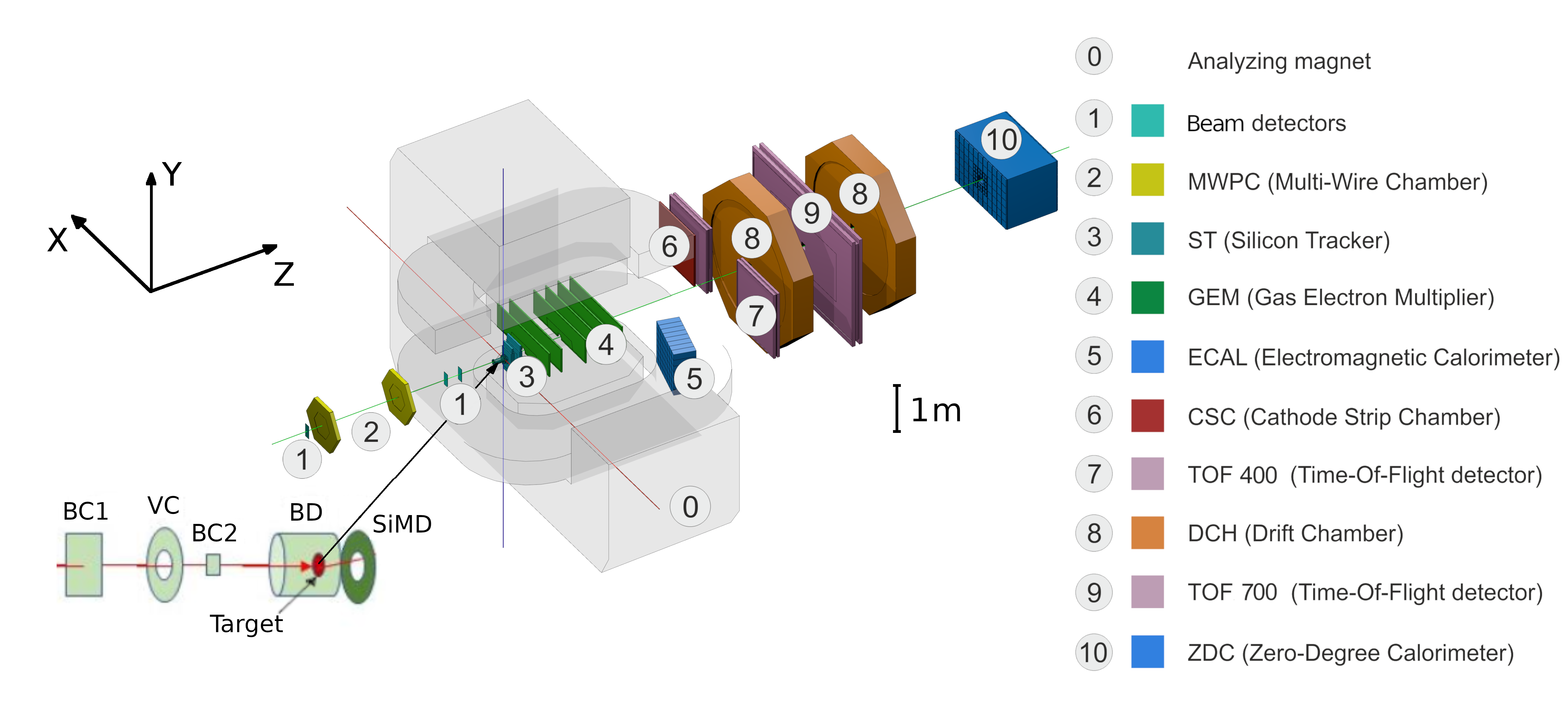}
\end{center}
\vspace{-0.9cm}
\caption {Schematic view of the BM$@$N setup in the argon beam run.}
\label{BMNsetup}
\end{figure}

The BM$@$N detector is a forward spectrometer covering the pseudorapidity
range of $1.6 \leq \eta \leq 4.4$. A schematic view of the BM$@$N setup in the
argon-beam run is shown in figure~\ref{BMNsetup}. A detailed description
of the setup is given in refs.~\cite{BMN_project,BMN_detectors}.
The spectrometer includes a central tracking system consisting of three planes
of forward silicon-strip detectors (ST) and six planes of detectors based on
gas electron multipliers (GEM)~\cite{BMN_GEM}. The central tracking system is
 located downstream of the target region inside of a dipole magnet
 with a bending power of about  2.1~Tm and with a gap of 1.05~m
 between the poles. In the measurements reported here, the central tracker
 covered only the upper half of the magnet acceptance.
\vspace{3.5cm}

Two sets of drift chambers (DCH), a cathode strip chamber (CSC), two sets
of time-of-flight detectors (ToF), and a zero-degree calorimeter (ZDC)
are located downstream of the dipole magnet.  The tracking system measures
the momenta of charged particles with a relative uncertainty
that varies from 2.5\% at 0.5 GeV/c to 2\% at 1--2 GeV/c and rises linearly to 6.5\% at 5 GeV/c.
The time resolutions of the time-of-flight systems ToF-400~\cite{BMN_ToF400_1,BMN_ToF400_2}  and
ToF-700~\cite{BMN_ToF700}  are 84~ps and 115~ps, respectively~\cite{KA_2022}.

 Two beam counters (BC1, BC2), a veto counter (VC), a barrel detector (BD),
 and a silicon multiplicity detector (SiMD) are used for event triggering
 and for the measurement of the incoming beam ions. The BC2 counter also provides
 the start time T0 for the time-of-flight measurements. The BD detector
 consists of 40 azimuthal scintillating strips arranged around the target,
 and the SiMD detector consists of 60 azimuthal silicon segments situated behind the target.

Data were collected with the argon beam with the intensity of a few 10$^5$ ions
per spill and a spill duration of 2--2.5~s. The kinetic energy of the beam was 3.2~A~GeV 
with a spread of about 1\%. A set of solid targets of various
materials (C, Al, Cu, Sn and Pb) with an interaction length of 3\% was used.
The following values of integrated luminosity were achieved for specific targets:  
2.1 {$\rm {\mu b^{-1}}$} (C), 2.3 {$\rm {\mu b^{-1}}$} (Al),
1.8 {$\rm {\mu b^{-1}}$} (Cu), 1.1 {$\rm {\mu b^{-1}}$} (Sn), 0.5 {$\rm {\mu b^{-1}}$} (Pb),
 with the total integrated luminosity of 7.8 {$\rm {\mu b^{-1}}$} obtained at the 
 end of data taking. A total of 16.3~M argon-nucleus collisions at 3.2~A~GeV 
 were reconstructed.

A logical beam trigger
BT = BC1$\land\overline{\rm{VC}}\land$BC2 was used to count the number of beam ions passing the target.
The following logic conditions were applied to generate the trigger signal: (1) BT$\land$(BD$\ge\rm{3,4}$);
(2) BT$\land$(SiMD$\ge\rm{3,4}$); (3) BT$\land$(BD$\ge\rm{2}$)$\land$(SiMD$\ge\rm{3}$).
The trigger conditions were varied to find the optimal ratio between the event
rate and the trigger efficiency for each target.
The trigger condition (1) was applied to 60\% of the data collected with the carbon target.
This trigger fraction was gradually decreasing with increasing the atomic weight of the target
down to 26\% for the Pb target. In contrast, the fraction of data collected with the trigger condition~(2) was increasing from 6\% for the carbon target up to 34\% for the Pb target.  The remaining data were collected with the trigger condition~(3).

\section{Event reconstruction}
\label{sect3}

Track reconstruction in the central tracker is based on a
``cellular automaton'' approach~\cite{Kisel2015,Kisel2006} implementing a constrained combinatorial search of
track candidates with their subsequent fitting by a Kalman filter to determine the track parameters.
These tracks are used to reconstruct primary and secondary vertices as well as global tracks by
extrapolation and matching to hits in the downstream detectors (CSC, DCH and ToF).

The primary collision vertex position (PV) is measured with a resolution of 2.4~mm in the X--Y plane
perpendicular to the beam direction and 3~mm in the beam direction.

Charged particles (protons, deuterons and tritons) are identified using the measured time of flight $\Delta t$ between T0 and the ToF detectors, the length of the trajectory $\Delta l$, and
the momentum $p$ reconstructed in the
central tracker. Then the squared mass $M^2$ of the particle is calculated by the
formula: $M^2 = p^2((\Delta t c/\Delta l)^2 - 1)$, where $c$ is the speed of light.

The following criteria are required for selecting proton, deuteron and triton candidates:
\begin{itemize}
\item Each track has at least four hits in the GEM detectors (six detectors in
total)~\cite{BMN_GEM}. Hits in the forward silicon detectors are used to
reconstruct the track, but no requirements are applied to the number of hits.

\item Tracks originate from the primary vertex. The deviation of the reconstructed
vertex $Z_{\mathrm {ver}}$ from the nominal target position along the beam direction  $Z_0$ is limited
to -3.4 cm $< Z_{\mathrm {ver}} - Z_0 <$ 1.7 cm.
The upper limit corresponds to  $\sim 5.7\sigma$ of the $Z_{ver}$ spread and cuts
off interactions with the trigger detector located 3~cm behind the target.
The beam interaction rate with the trigger detector itself is well below 1\%, and it was neglected in Monte Carlo modeling of the experimental setup because its contribution  was estimated within the modeling uncertainties.

\item Distance of closest approach (DCA) of the track to the primary vertex in the X--Y plane at Z$_{\mathrm{ver}}$
is required to be less than 1~cm, which corresponds to 4$\sigma$ of the vertex residual distribution in the X--Y plane.

\item Momentum range of positively charged particles 
is limited by the acceptance of the ToF-400 and ToF-700 detectors to $p>0.5$ GeV/c and $p>0.7$~GeV/c, respectively.

\item Distance of extrapolated tracks to the CSC (DCH) hits as well as to the
ToF-400 (ToF-700) hits should be within $\pm 2.5\sigma$ of the momentum dependent
hit-track residual distributions.

\end{itemize}

\begin{figure}[tbh]
\begin{center}
\vspace{-0.2cm}

\hspace{0.5cm} (a) \hspace{7cm} (b)
\includegraphics[width=0.49\textwidth,bb=0 0 1700 997]{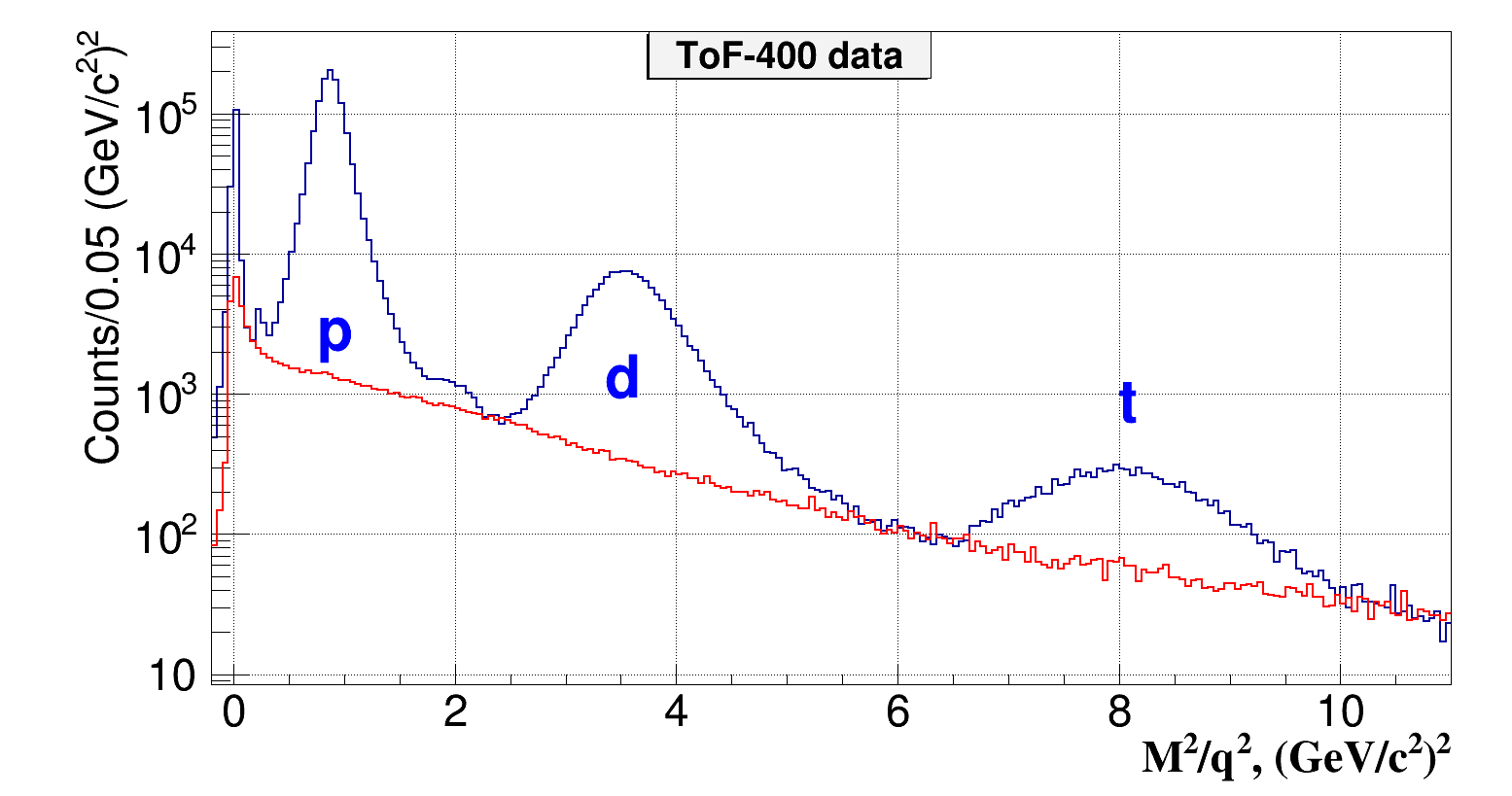}
\includegraphics[width=0.49\textwidth,bb=0 0 1700 997]{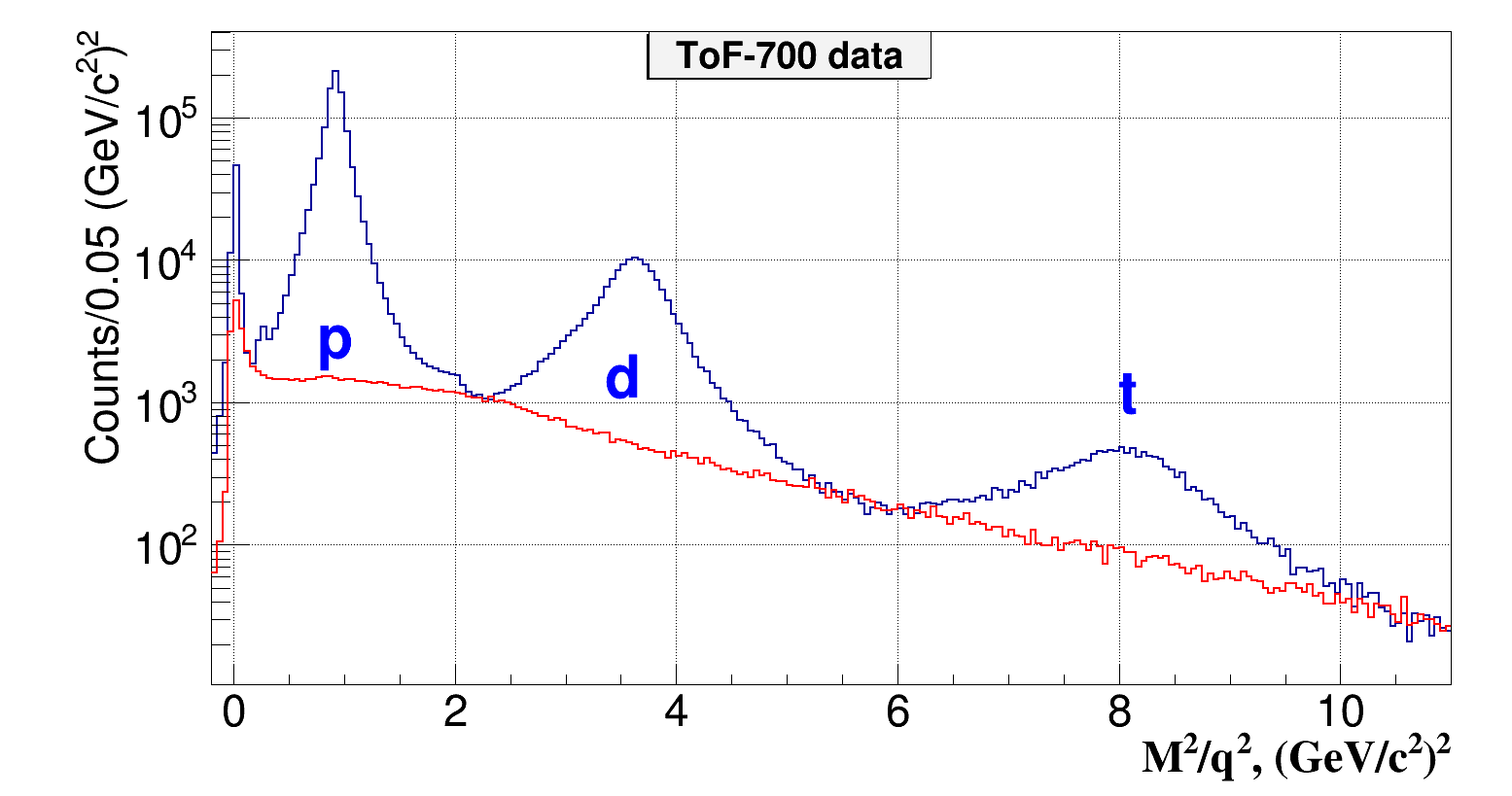}
 \end{center}
 \vspace{-0.7cm}
 \caption{$M^2/q^2$ spectra of positively charged particles produced in argon-nucleus
 interactions and measured in the ToF-400 (a) and ToF-700 (b) detectors.
 Peaks of protons, deuterons and tritons with the charge $q=1$ are indicated; 
 the small peaks of He fragments with $q=2$ either overlap with the deuteron 
 peaks ($^4$He) or show up at $M^2/q^2 \sim 2$ (GeV/$c^2)^2$ ($^3$He). The background 
 estimated from ``mixed events'' is shown by the red line histograms.}
 \label{m2tof400and700}
\end{figure}

The mass squared ($M^2$) spectra of positively charged particles produced
in interactions of the 3.2~A~GeV argon beam with various targets are shown in
figures~\ref{m2tof400and700}a and \ref{m2tof400and700}b for ToF-400 and ToF-700 data,
respectively. Particles that satisfy the above selection criteria contribute to 
the $M^2$ spectra. The proton, deuteron and triton signals are extracted in $M^2$ 
windows, which depend on rapidity, and extend
within 0.4--1.7 (GeV/c$^2)^2$, 2.3--5.0 (GeV/c$^2)^2$ and 6.6--10.0 (GeV/c$^2)^2$, 
at the maximal rapidity, respectively. The signals of protons,  deuterons and tritons 
and their statistical errors are calculated as : $sig=hist-bg$, 
where $hist$ denotes the histogram integral yield within the selected 
$M^2$-window, and $bg$ is the background.

The shape of the background under the proton, deuteron and triton signals in the $M^2$ spectra is
estimated using the ``mixed event'' method. For that, tracks reconstructed in the central
tracker are matched to hits in the ToF detectors taken from different events containing 
a similar number of tracks. The ``mixed event''
background is normalized to the integral of the signal histogram outside the $M^2$
windows of protons, deuterons and tritons. It is found that the background level
differs for light and heavy targets and for different intervals of rapidity
and transverse momentum.

\begin{figure}[htb]
\begin{center}
\vspace{-0.1cm}

\hspace{0.5cm} (a) \hspace{5.5cm} (b)
\includegraphics[width=0.48\textwidth,bb=0 0 1700 997]{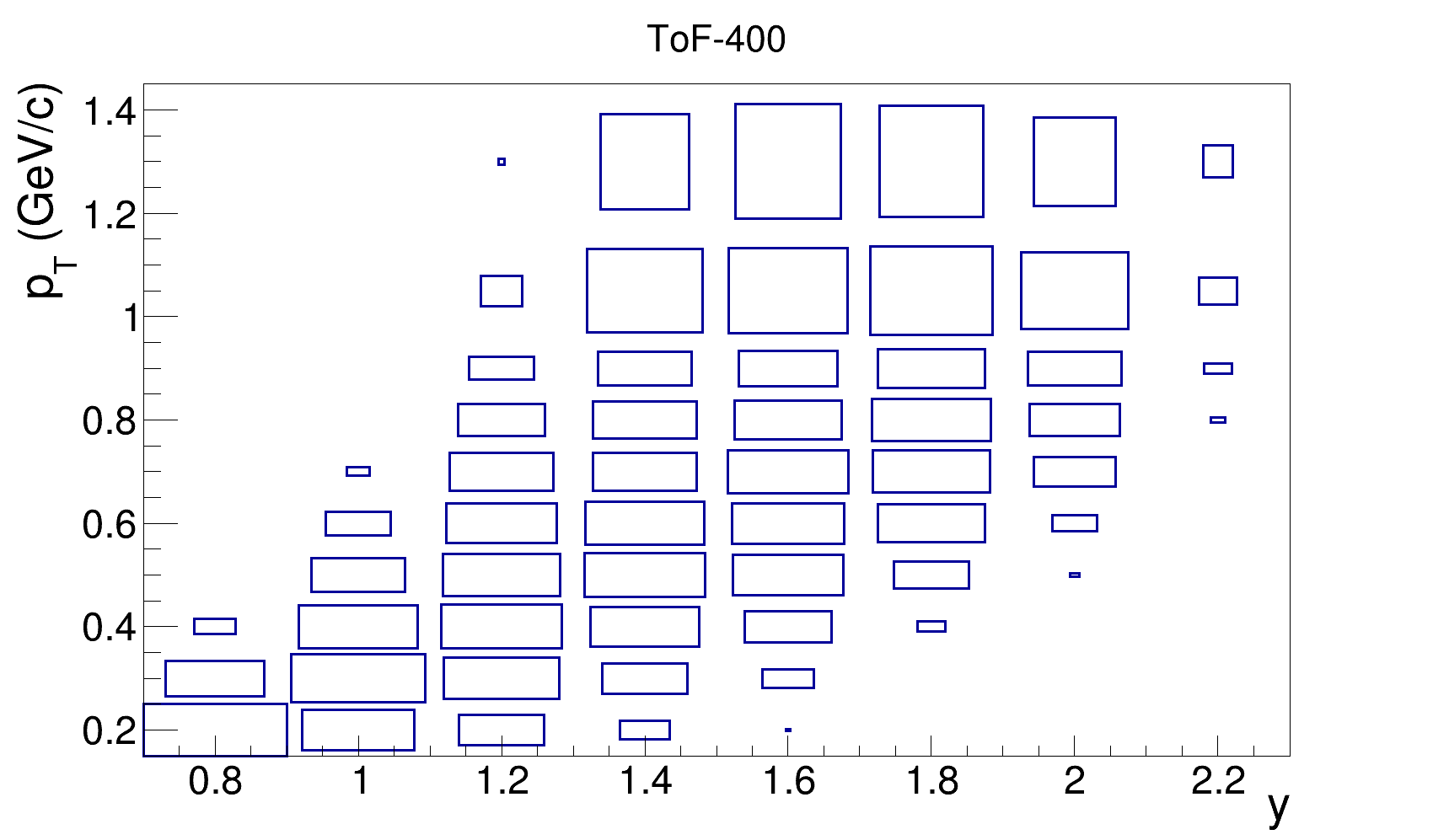}
\includegraphics[width=0.48\textwidth,bb=0 0 1700 997]{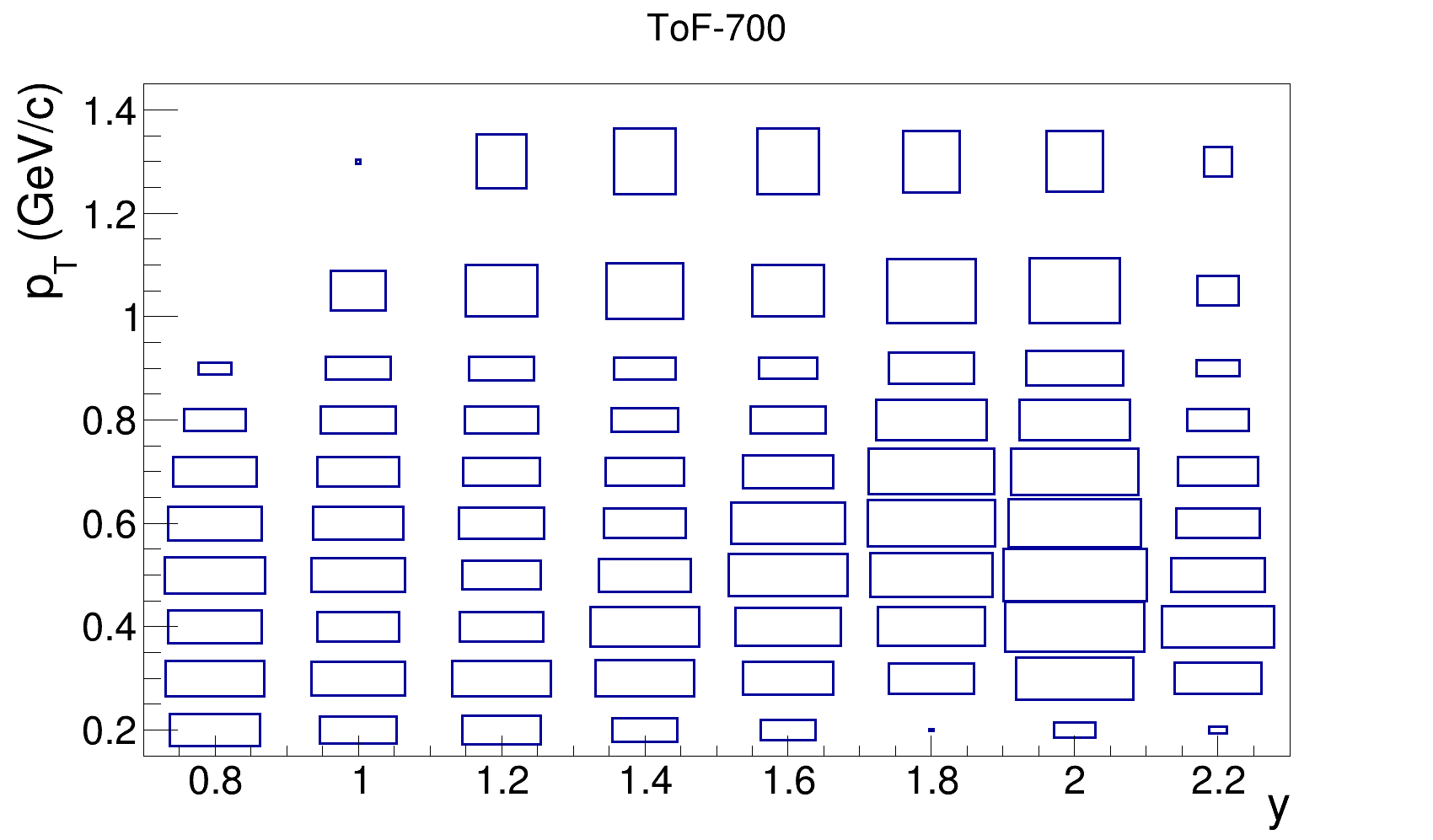}
 \end{center}
 \vspace{-1.0cm}
\caption{Distribution of the deuteron signals measured in ToF-400 (a)
          and ToF-700 (b) in the rapidity vs. transverse momentum plane in Ar+Sn interactions.}
 \label{ypt_pdt_sn}
\end{figure}

\begin{figure}[htb]
\begin{center}
\vspace{-0.1cm}

\includegraphics[width=0.7\textwidth]{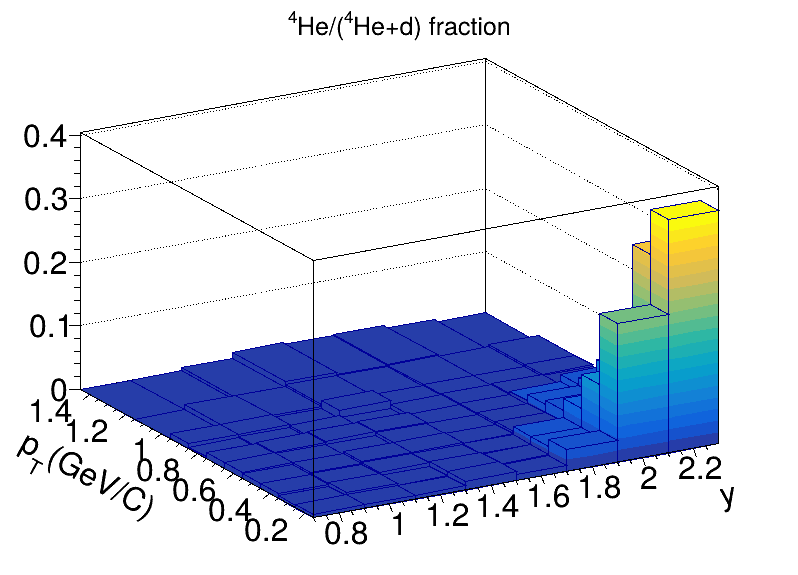}
 \end{center}
 \vspace{-1.0cm}
\caption{Fraction of $^4$He in  the $^4$He+$d$ sample measured in the rapidity vs. transverse momentum
           plane in Ar+A interactions.}
 \label{H4fract}
\end{figure}

The ToF-400  and ToF-700  detectors cover different ranges of rapidity and transverse
momentum of detected particles.  The
deuteron signals from Ar+Sn interactions measured by ToF-400 and ToF-700  are shown 
in figure~\ref{ypt_pdt_sn} in the rapidity vs. transverse
momentum plane before making any efficiency corrections.

The $dE/dx$ information from  the GEM detectors is used to separate the deu\-teron signals 
from the overlapping TOF $^4$He signals. The fraction of $^4$He in the entire $^4$He+$d$ 
sample is determined in rapidity and transverse momentum
 bins and subtracted from the deuteron TOF signals. The $^4$He fraction combined for all 
 the targets is presented in figure~\ref{H4fract}. As can be seen, in most of the $y-p_T$ bins, 
 the $^4$He fraction is below 3\%. However, it reaches 20--35\% in a few bins at large $y$ 
 and low $p_T$, associated with spectator $d$ and $^4$He, with a large fraction of $^4$He.

\section{Reconstruction efficiency  and trigger performance}
\label{sect4}

In order to evaluate the proton, deuteron and triton reconstruction efficiency, Monte Carlo data
samples of argon-nucleus collisions were produced with the DCM-SMM~\cite{DCM_QGSM,DCM_SMM} event
 generator. The propagation of particles through the entire
 detector volume and responses of the detectors were simulated using the
 GEANT3 toolkit~\cite{GEANT3} integrated into the BmnRoot software framework~\cite{BmnRoot}. 

The Monte Carlo events passed through the same chain of reconstruction and identification
as the experimental ones. 
The efficiencies of the silicon, GEM, CSC, DCH and ToF detectors were
adjusted in the simulation in accordance with  the measured detector efficiencies~\cite{DetEff}.
More details of the simulation are given in ref.~\cite{BMN_piKpaper}.

\begin{figure}[!htb]
\begin{center}
\hspace{0cm} (a) \hspace{5.0cm} (b)

\includegraphics[width=0.43\textwidth]{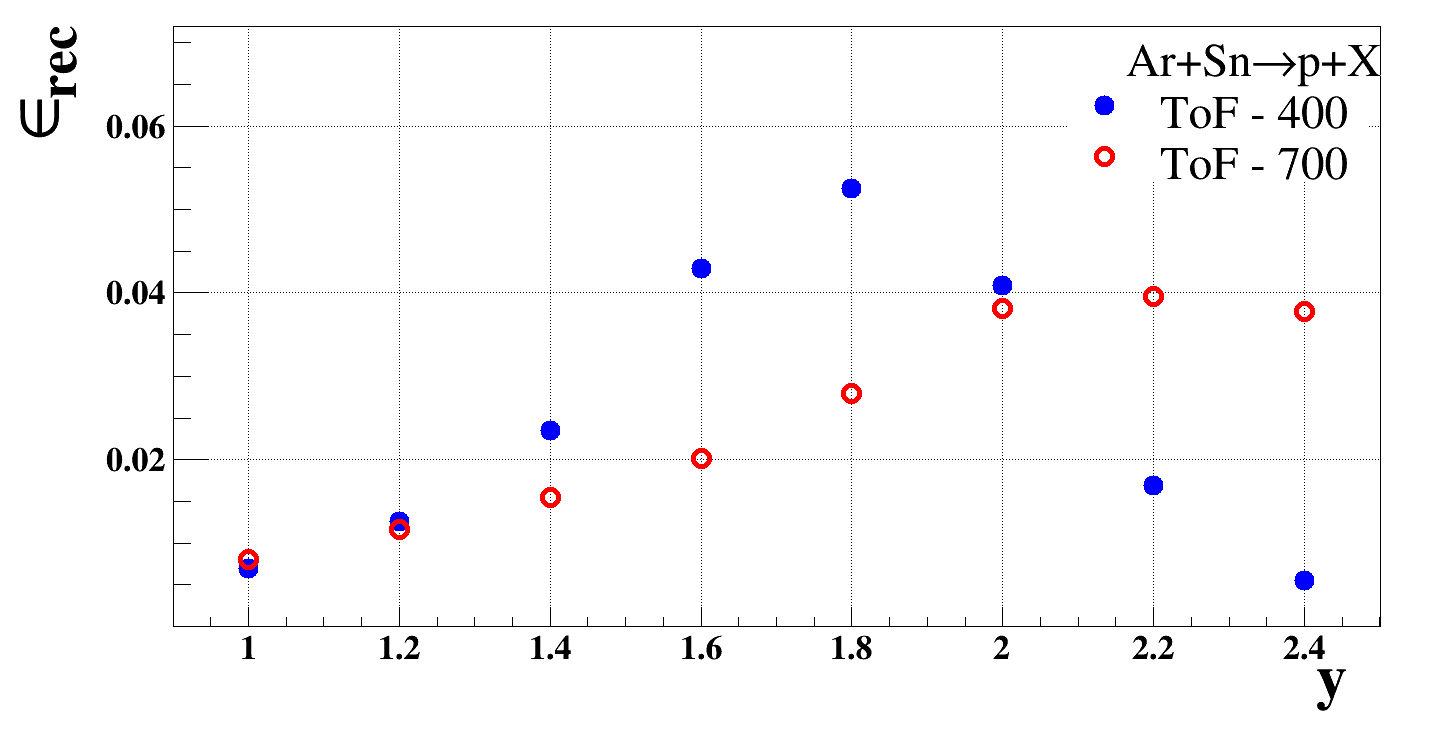}
\includegraphics[width=0.45\textwidth]{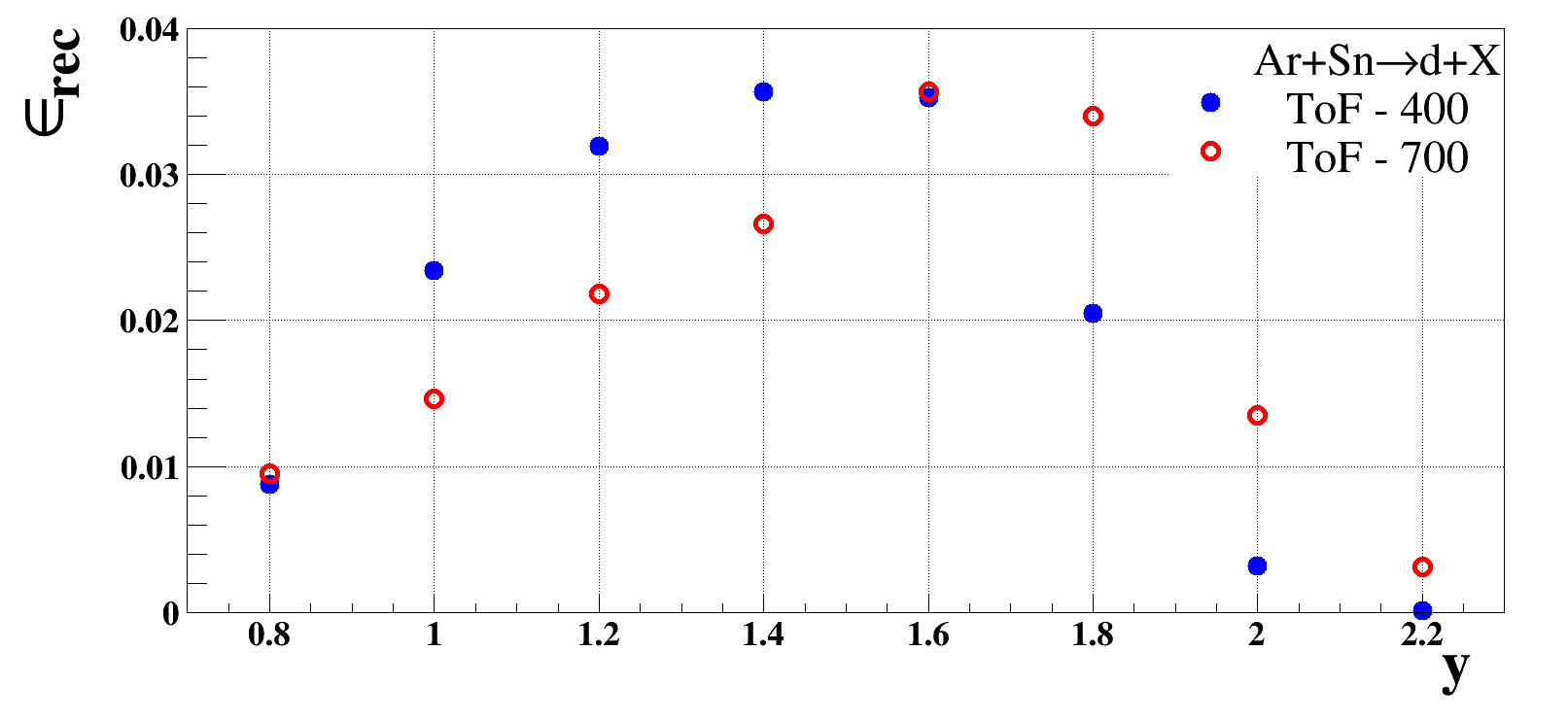}
\includegraphics[width=0.45\textwidth]{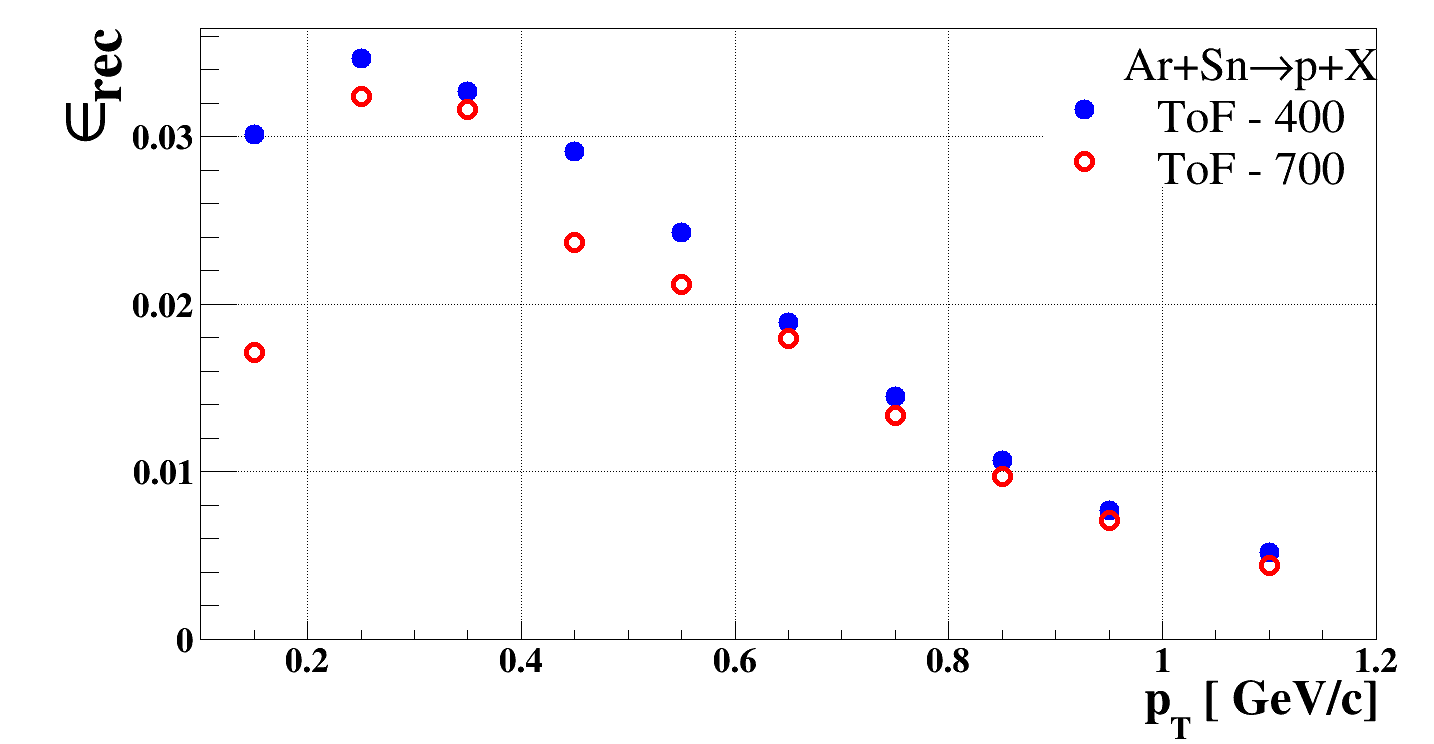}
\includegraphics[width=0.45\textwidth]{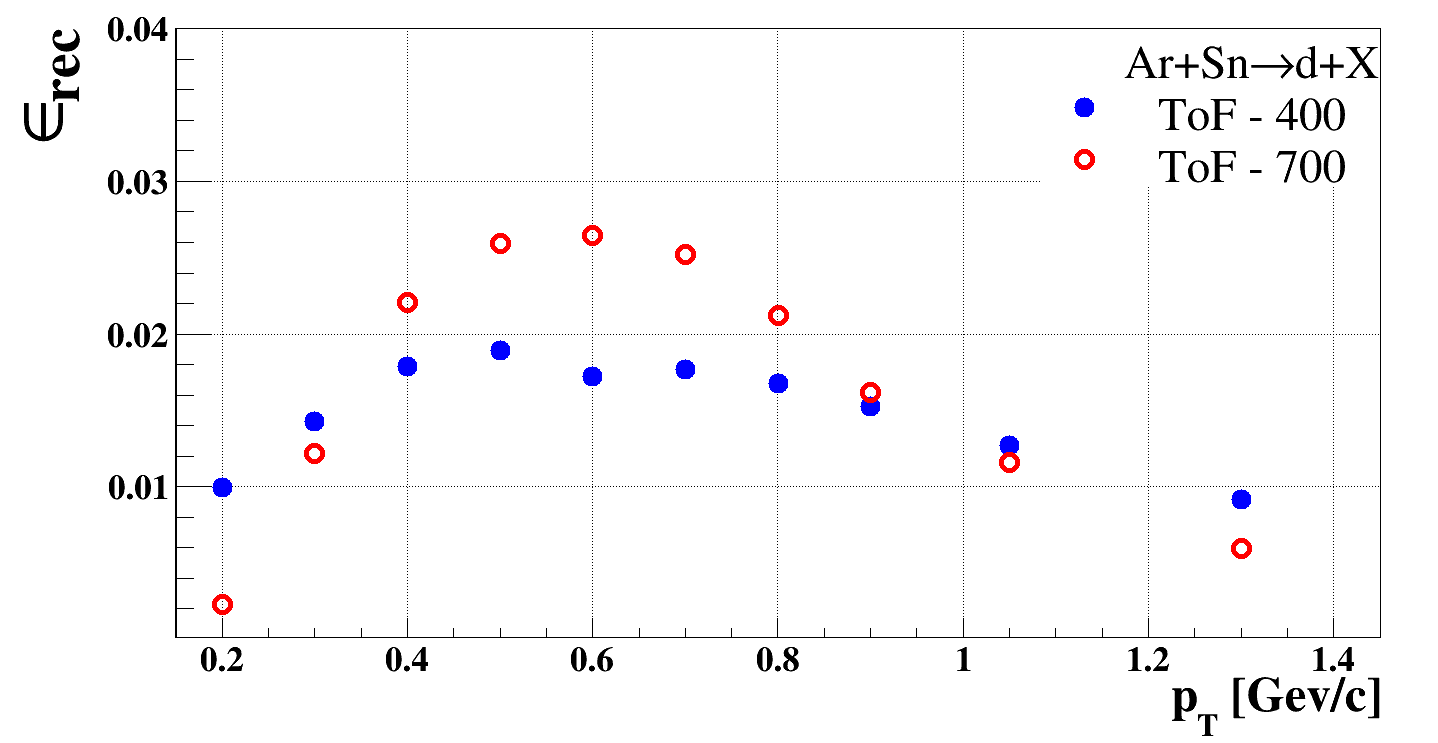}
 \end{center}
 \vspace{-0.8cm}
 \caption{Reconstruction efficiency of protons (a) and deuterons (b) produced in Ar+Sn collisions,
 detected in ToF-400 (full blue circles) and ToF-700 (open red circles)
 as functions of  rapidity $y$ and  $p_T$, see text for details.}
 \label{eff}
\end{figure}

The proton, deuteron and triton reconstruction efficiencies are calculated in
intervals of rapidity $y$ and transverse momentum $p_T$. The reconstruction
efficiency includes geometrical acceptance, detector efficiency,
kinematic and spatial cuts, and the loss of protons, deuterons and tritons due to 
in-flight interactions.
The resulting reconstruction efficiencies in ToF-400 and ToF-700 are shown in 
figure~\ref{eff} for protons (left) and deuterons (right) from Ar+Sn interactions 
as functions of $y$ (upper panels) and $p_T$ (lower panels).

The trigger efficiency $\epsilon_{trig}$ depends on the number of fired channels
in the BD (SiMD) detectors. It was calculated for events with reconstructed protons,
deuterons and tritons using event samples recorded with an independent trigger based
on the SiMD (BD) detectors. The BD and SiMD detectors cover different and
non-overlapping regions of the BM$@$N acceptance, that is, they detect different
collision products.

The efficiency of the combined BD and SiMD triggers was calculated as the product of
the efficiencies of the BD and SiMD triggers.
The trigger efficiency decreases with the decrease of the target mass and with 
the increase of the collision centrality. More details on the 
evaluation of the trigger efficiencies  are given in ref.~\cite{BMN_piKpaper}. 
In particular, as illustrated in figure 10 of \cite{BMN_piKpaper}, the trigger 
system accepts events in the full centrality range.

\section{Centrality classes}
\label{sect44}

\begin{figure}[tbh]
 \begin{center}
\vspace{-0.5cm}
\hspace{0cm} (a) \hspace{5cm} (b)

\includegraphics[width=0.48\textwidth]{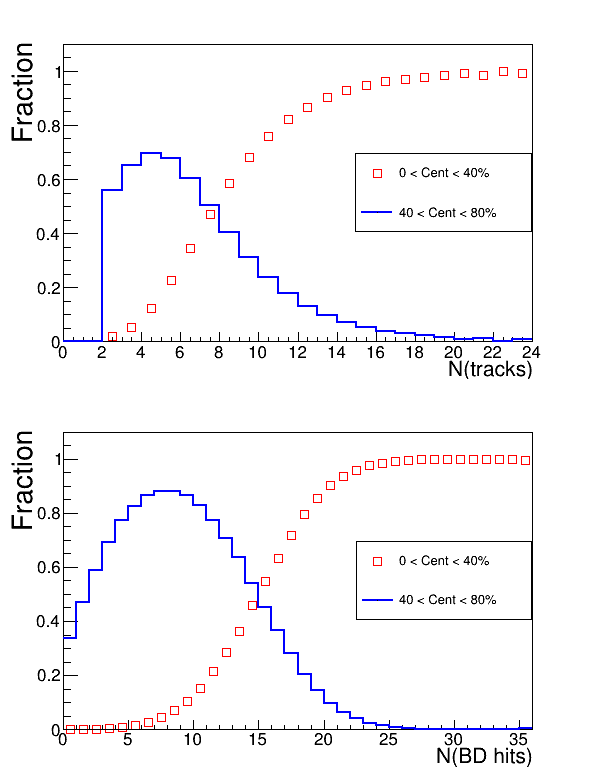}
\includegraphics[width=0.49\textwidth]{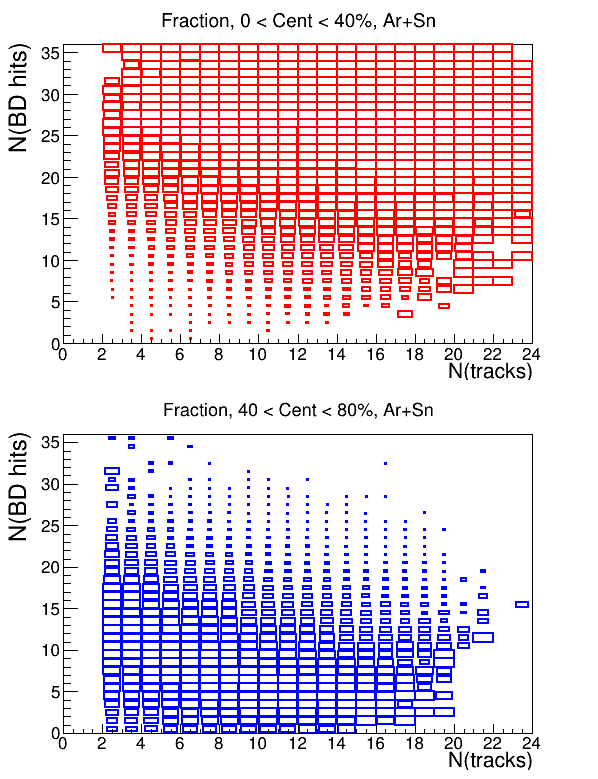}

\end{center}
\vspace{-0.8cm}
 \caption{(a) Probability distribution of the number of tracks N(tracks) in the 
 primary vertex (upper panel) and the number of hits N(BD) in the BD detector 
 (lower panel) for events with centrality 0--40\% (red open symbols) and 
 40--80\% (blue histogram); (b) Two-dimensional plot of the probability distribution 
 of  N(tracks) (horizontal axis) vs. N(BD) (vertical axis) in events with 
 centrality 0--40\% (upper panel) and 40--80\% (lower panel).
 }
 \label{Ntr_Nbd_centr}
\end{figure}

The  nucleus-nucleus collision centrality is defined as the ratio  of
the interaction cross section corresponding to a given impact parameter interval $[0,b]$ 
to the total inelastic interaction cross section.
Two classes of centrality:  0--40\% (more central collisions) and  40--80\% (more peripheral 
collisions) are defined from the impact parameter distributions of Ar+A inelastic interactions
 simulated by the DCM-SMM model. The boundary impact parameters $b_{40}$ and $b_{80}$ for the 
 definition of the two classes for interactions of Ar with various targets are given 
 in table~\ref{table_bcross}.
It was found that the number of tracks originating from the primary event vertex N(tracks)
 and the number of hits in the Barrel Detector N(BD) are anti-correlated with the impact parameter~$b$.
Using the results of the DCM-SMM Monte Carlo simulation, the fractions of reconstructed events, 
which belong to the centrality classes 0--40\% and 40--80\%,  are calculated. Fractions of events 
with centrality 0--40\% and 40--80\% are presented in
figure~\ref{Ntr_Nbd_centr} as functions of N(tracks), N(BD) and as a two-dimensional 
distribution N(tracks) / N(BD).

\begin{table}[hbp]

  \caption {Boundary impact parameters $b_{40}$ and $b_{80}$ for the definition of the centrality classes 
  0--40\% and 40--80\% and the inclusive inelastic cross section $\sigma_{inel}$~\cite{HadesL0} for Ar+A interactions.}
\begin{footnotesize}
\vspace{0.5cm}

\begin{tabular}{|l|c|c|c|c|c|}
\hline
& & & & & \\
& Ar+C & Ar+Al & Ar+Cu & Ar+Sn & Ar+Pb \\
& & & & & \\
\hline
& & & & & \\
$b_{40}$, fm                               & $4.23$  & $4.86$ & $5.66$ & $6.32$  & $7.10$  \\
& & & & & \\
$b_{80}$, fm                               & $6.2$  & $7.0$ & $8.0$ & $9.0$  & $10.0$  \\
& & & & & \\
$\sigma_{inel}$, mb  \cite{HadesL0} & $1470 \pm 50$  & $1860 \pm 50$ & $2480 \pm 50$ & $3140 \pm 50$  & $3940 \pm 50$  \\
& & & & & \\
\hline
\end{tabular}
\end{footnotesize}
\label{table_bcross}
\end{table}

Fractions (probabilities) of events with centrality 0--40\% and 40--80\%, taken from the 
two-dimensional N(tracks)/N(BD) distributions, are used as event weights to define the 
weighted numbers of reconstructed protons, deuterons and tritons
in the $y$ and $p_T$ bins in data as well as in simulation.
The systematic uncertainty of the event centrality is estimated from the remaining 
difference between the simulations and data in the shape of the N(tracks) and N(BD) 
distributions in $y$ and $p_T$ bins.

\section{Cross sections, multiplicities and systematic uncertainties}
\label{sect5}

The protons, deuterons and tritons from  interactions of Ar with C, Al, Cu, Sn and Pb  are measured in the
following kinematic ranges: transverse momentum $0.1<p_T<1.2$~GeV/c (protons), $0.15<p_T<1.45$~GeV/c (deuterons),
$0.2<p_T<1.6$~GeV/c (tritons) and rapidity in the laboratory frame $0.9<y<2.5$ (protons), 
$0.7 <y<2.3$ (deuterons), $0.7 <y<2.1$ (tritons).
The differential cross sections {\small $d^2\sigma_{p,d,t}(y,p_T)/dydp_T$} and
multiplicities {\small $d^2N_{p,d,t}(y,p_T)/dydp_T$} of protons, deuterons and tritons
produced in Ar+C, Al, Cu, Sn and Pb interactions are calculated using the relations:

\begin{footnotesize}
\vspace{0.3cm}
$d^2\sigma_{p,d,t}(y,p_T)/dydp_T =  \Sigma [ d^2 n_{p,d,t}(y,p_T, N_{tr}) / 
(\epsilon_{trig}(N_{tr}) dy dp_T)] \times 1 / ( L \epsilon_{p,d,t}^{rec}(y,p_T))$
\begin{equation}\tag{1}
 d^2N_{p,d,t}(y,p_T)/dydp_T = d^2\sigma_{p,d,t}(y,p_T) / (\sigma_{inel} dydp_T),
\end{equation}
\end{footnotesize}

\noindent where the sum is performed over bins of the number of tracks in the
primary vertex $N_{tr}$; $n_{p,d,t}(y, p_T, N_{tr})$ is the number of
reconstructed protons, deuterons and tritons in the intervals $dy$ and $dp_T$;
$\epsilon_{trig}(N_{tr}) $ is the track-dependent trigger efficiency;
$\epsilon_{p,d,t}^{rec}(y,p_T)$ is the reconstruction efficiency of protons, deuterons and tritons; $L$
is the luminosity;   and $\sigma_{inel}$ is the inelastic cross section for argon-nucleus interactions.
The cross sections and multiplicities  are  evaluated for the two centrality classes: 0--40\% and 40--80\%.
\begin{table}[!hbp]
\vspace{-0.5cm}

\caption{
         {Mean relative systematic uncertainties  (in \%) averaged over the $y$, $p_T$ 
	 ranges of protons, deuterons and tritons
	    measured in argon-nucleus interactions.}}
\begin{footnotesize}
\vspace{-0.3cm}
\hspace{2.5cm}
\begin{center}
\begin{tabular}{|l|c|c|c|c|c|}
\hline
 & Ar+C & Ar+Al & Ar+Cu & Ar+Sn & Ar+Pb \\
\hline
 & & & & & \\
$\epsilon_{trig}$ $p,d,t$    &   9  &  7  &  7  &  7   &  7 \\
 & & & & & \\
 protons                   &      &     &     &      &    \\
$n_p/\epsilon_{rec}$   &  15  &   6 &   8 &  14  &  11 \\
 Total                     &  18  &   9 &  11 &  16  &  13 \\
 & & & & & \\
deuterons                  &      &     &     &      &    \\
 $n_d/\epsilon_{rec}$  &   32 &  22 &  20 &  19  &  22 \\
Total                      &   33 &  23 &  21 &  20  &  23 \\
 & & & & & \\
tritons                    &      &     &     &      &   \\
 $n_t/\epsilon_{rec}$  &   43 &  22 &  20 &  20  &  22 \\
Total                      &   44 &  23 &  21 &  21  &  23 \\
\hline
\end{tabular}
\end{center}
\end{footnotesize}
\label{uncertainties_tablepiK}
\end{table}

Several sources of systematic uncertainties  are considered in evaluating the uncertainties 
of the measured proton, deuteron and triton yields $n_{p,d,t}$ and the reconstruction
efficiency $\epsilon_{rec}$. 
Some of them affect both the yield $n_{p,d,t}$ and the reconstruction
efficiency $\epsilon_{rec}$. For these cases, the impact of correlations between them 
on the $n_{p,d,t}/\epsilon_{rec}$ ratio is taken into account.
The systematic uncertainties associated with the track reconstruction as well as with 
the trigger efficiency are discussed in detail in ref.~\cite{BMN_piKpaper}. Additional 
sources specific to this analysis are listed below: 
\begin{itemize}
\vspace{-0.3cm}
\item Systematic uncertainty of the background subtraction in the mass-squared $M^2$ 
spectra of identified particles: it is estimated as the difference between the background 
integral under the $p, d, t$ mass-squared windows taken from ``mixed events'' 
(as described in section~\ref{sect3}) and from the fitting of the $M^2$ spectra by a 
linear function.  The latter is done in the $M^2$ range,
 excluding the proton, deuteron and triton signal windows.
\vspace{-0.3cm}
\item  Systematic uncertainty calculated as half of the difference between the p/d/t 
       yield measured in the ToF-400 and ToF-700 detectors in bins of rapidity $y$. 
\vspace{-0.3cm}
\item Systematic uncertainty of the event centrality weights estimated (i) from 
      the remaining difference in the shape of the N(track) and N(BD)
      distributions in  $y$ and $p_T$ bins in the data and the simulation; (ii) from the 
      difference in the event centrality  weights taken from the two-dimensional  
      N(track)/N(BD) distribution relative to the one-dimensional N(BD) distribution.
\vspace{-0.3cm}
\end{itemize}
\noindent Table~\ref{uncertainties_tablepiK} summarizes the mean values (averaged over $p_T$, $y$ and
$N_{tr}$)
of the systematic uncertainties of the various factors of eq.~(1), $n_{p,d,t}$, $\epsilon_{rec}$,
and $\epsilon_{trig}$. 
The total systematic uncertainty from these sources,
calculated as the square sum of their uncertainties from different sources,
is listed in table~\ref{uncertainties_tablepiK} for each target.

The luminosity is calculated from the beam flux $\Phi$ as given by the beam trigger
(see section \ref{sect2}) and the target thickness $l$ using the relation $ L = \Phi \rho l$, where $\rho $
is the target density expressed in atoms/cm$^3$. The systematic uncertainty of the luminosity
is estimated from the fraction of the beam that can miss the target, determined from the
vertex positions, and found to be within 2\%.
The inelastic cross sections of Ar+C, Al, Cu, Sn and Pb interactions are taken from the predictions of the DCM-SMM model. 
The $\sigma_{inel}$ uncertainties for Ar+C, Al, Cu, Sn and Pb interactions  given in 
table~\ref{table_bcross} are estimated from the empirical formulae taken from ref. \cite{HadesL0,AngelovCC}.

\section{Rapidity and transverse mass spectra}
\label{sect6}

 At a kinetic energy of 3.2~A~GeV,
 the rapidity of the nucleon-nucleon center-of-mass (CM) system is $y_{CM}=1.08$. The rapidity
 intervals covered in the present measurements, $0.9<y<2.5$, $0.7<y<2.3$ and $0.7<y<2.1$ for  protons, deuterons and tritons,
  respectively, correspond, therefore, to the forward and central rapidity regions in the nucleon-nucleon
 CM system. The measured yields of protons, deuterons and tritons in $m_T$ and $y$ bins in the 
 two centrality intervals in Ar+C, Al, Cu, Sn and Pb interactions can be found in ref.~\cite{BMN_yields}.

\begin{figure}[htpb]
\begin{center}
\includegraphics[width=0.95\textwidth,bb=0 0 1266 607]{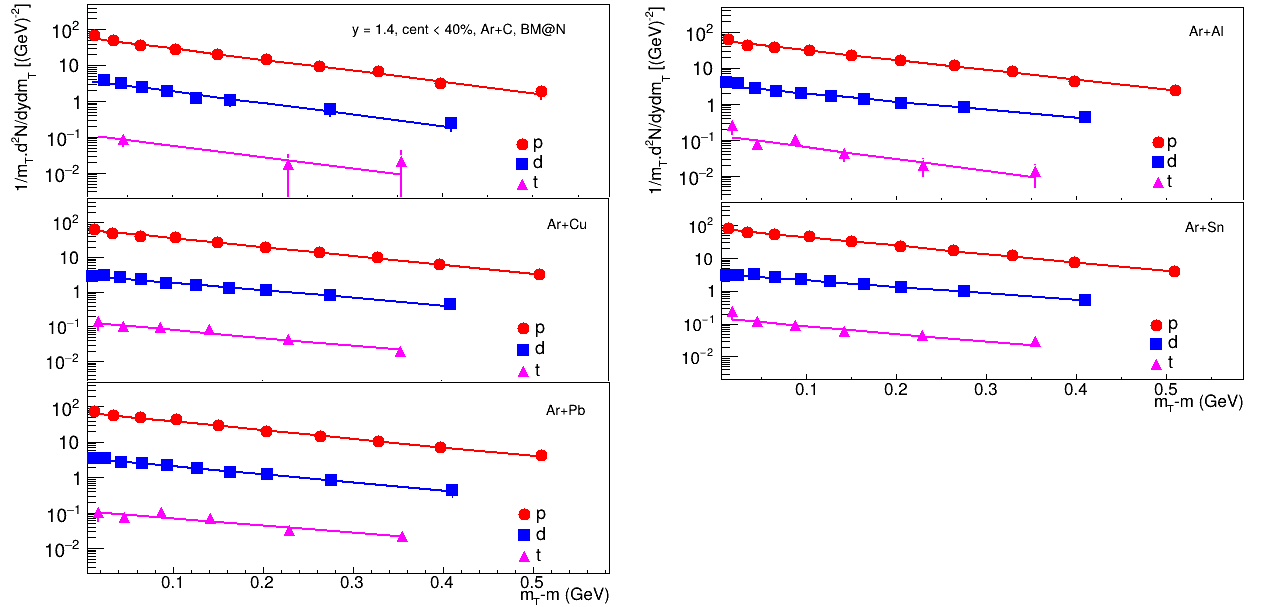}
\end{center}
\vspace{-0.8cm}

 \caption{Invariant transverse mass spectra of protons, deuterons and tritons produced at rapidity $y=1.4$ 
 in  Ar+C, Al, Cu, Sn and Pb
 interactions  with centrality  0--40\%. The vertical bars and boxes represent the statistical
 and systematic uncertainties, respectively. The lines show the results of the fit by an exponential function.}
 \label{mtyields}
\end{figure}

As an example, figure~\ref{mtyields} shows the invariant transverse mass $m_T=\sqrt{m^2+p_T^2}$ spectra 
of protons, deuterons and tritons ($m=m_{p,d,t}$) produced at $y=1.4$ in Ar+A collisions in the 0--40\% centrality class.
The spectra are parameterized by an exponential function as:
\begin{equation}\tag{2}
 \frac{1}{m_T} d^2N/dydm_T = \frac{dN/dy}{T_0(T_0+m)} {\rm exp}(-(m_T-m)/T_0), 
\label{pTdep}
\end{equation}
\noindent where the fitting parameters are the integral of the $m_T$ spectrum, $dN/dy$,
and the inverse slope, $T_0$. The $dN/dy$ and $T_0$ values extracted from the fit can be found in ref.~\cite{BMN_yields}. 

\begin{figure}[htpb]
\begin{center}
\vspace{-0.5cm}

\hspace{0cm} (a)

\includegraphics[width=0.8\textwidth,bb=0 0 748 475]{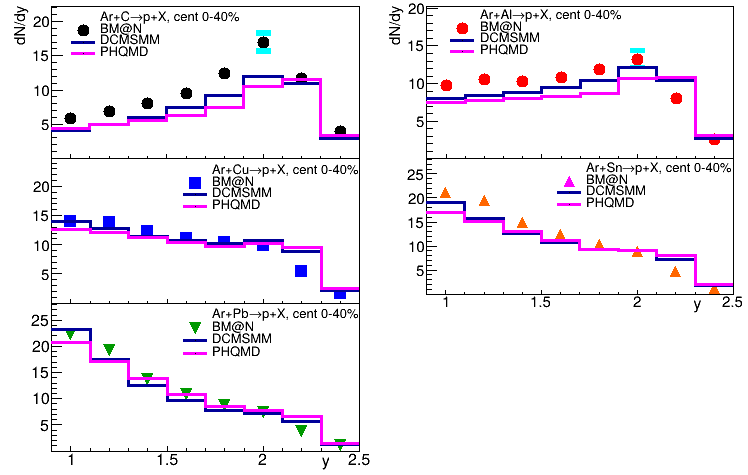}
\hspace{3cm} (b)

\includegraphics[width=0.75\textwidth,bb=0 0 748 475]{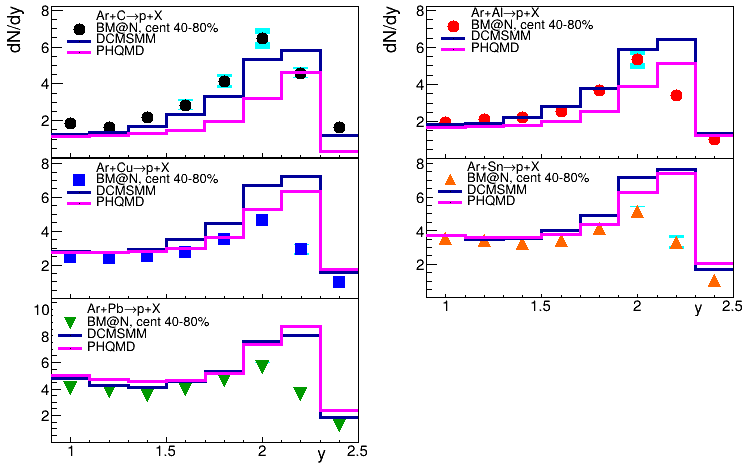}
\end{center}
\vspace{-0.8cm}

 \caption{Rapidity distributions $dN/dy$ of protons produced in  Ar+C, Al, Cu, Sn and Pb
 interactions at 3.2~A~GeV with centrality  0--40\% (a) and 40--80\% (b). The results are integrated 
 over $p_T$. The vertical bars and boxes represent the statistical
 and systematic uncertainties, respectively. The predictions of the DCM-SMM and PHQMD models 
 are shown as blue and magenta histograms.}
 \label{yields_yp}
\end{figure}

\begin{figure}[htpb]
\begin{center}
\vspace{-0.5cm}

\hspace{0cm} (a)

\includegraphics[width=0.8\textwidth,bb=0 0 748 475]{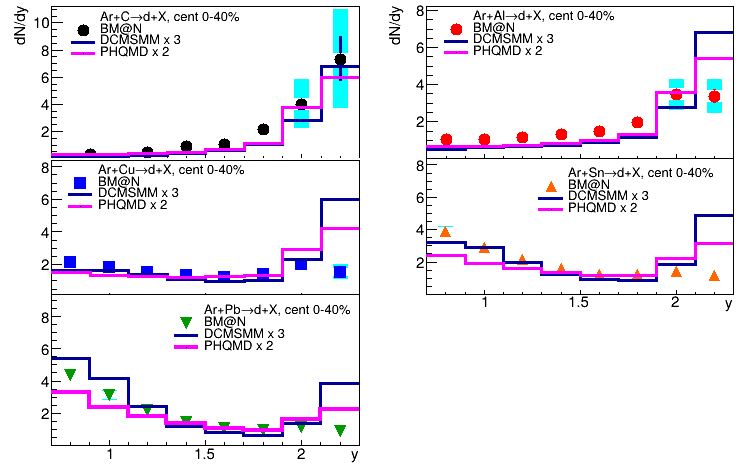}
\hspace{3cm} (b)

\includegraphics[width=0.8\textwidth,bb=0 0 748 475]{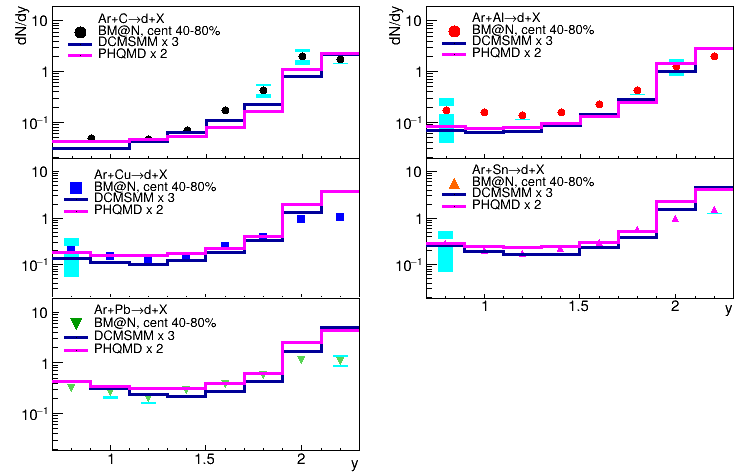}
\end{center}
\vspace{-0.8cm}

 \caption {Rapidity distributions $dN/dy$ of deuterons  produced in  Ar+C, Al, Cu, Sn and Pb interactions
 with centrality 0--40\% (a) and 40--80\% (b). The results are integrated over $p_T$.
 The vertical bars and boxes represent the statistical and systematic uncertainties,
 respectively. The predictions of the DCM-SMM and PHQMD models, multiplied by factors of 3 and 2, 
 respectively, are shown as blue  and magenta histograms.}
 \label{yields_yd}
\end{figure}

\begin{figure}[htpb]
\begin{center}
\vspace{-0.5cm}
\hspace{0cm} (a)

\includegraphics[width=0.8\textwidth,bb=0 0 748 475]{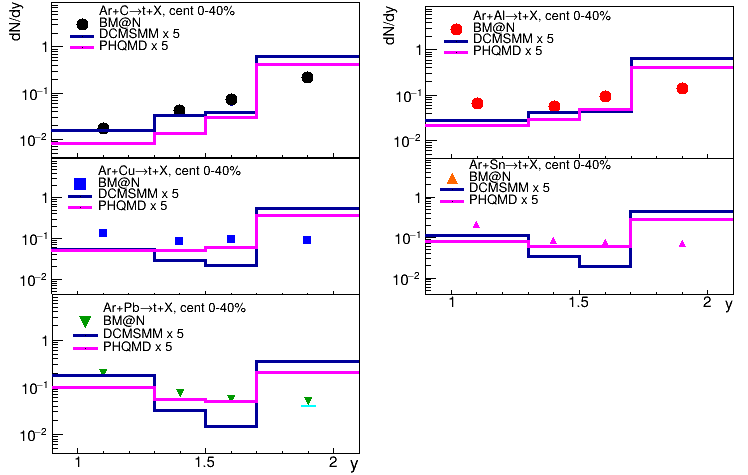}
\hspace{3cm} (b)

\includegraphics[width=0.8\textwidth,bb=0 0 748 475]{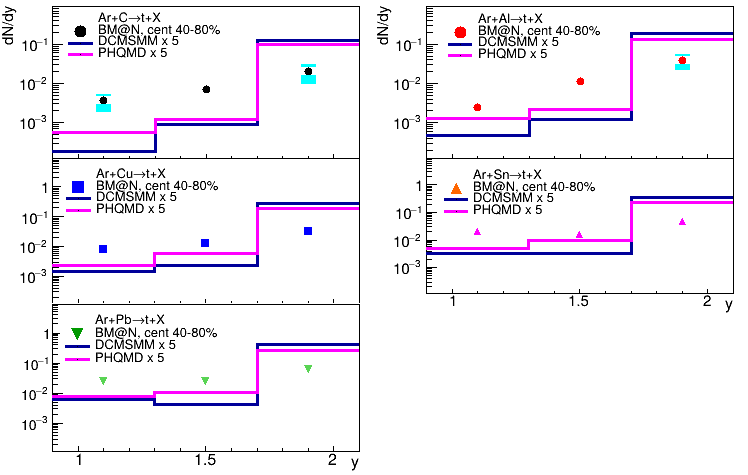}
\end{center}
\vspace{-0.8cm}
 \caption {Rapidity distributions $dN/dy$ of tritons  produced in  Ar+C, Al, Cu, Sn and Pb interactions
 with centrality 0--40\% (a) and 40--80\% (b). The results are integrated over $p_T$.
 The vertical bars and boxes represent the statistical and systematic uncertainties,
 respectively. The predictions of the DCM-SMM and PQHMD models, multiplied by a factor of 5, are shown 
 as blue and magenta histograms.}
 \label{yields_yt}
\end{figure}

The $dN/dy$ distributions of protons, deuterons and tritons produced in Ar+A collisions with 
centrality 0--40\%  are shown in figures~\ref{yields_yp}(a), \ref{yields_yd}(a) and~\ref{yields_yt}(a). 
The comparison of the measurements  with the predictions of the DCM-SMM and PHQMD models is also shown 
in these figures. The boundary impact parameters $b_{40}$  and $b_{80}$ listed in table~\ref{table_bcross} 
are used to define the centrality classes in the model calculations. 

As can be seen, the shapes of the rapidity distributions  of $p, d, t$ essentially vary with the target mass.
For protons, the predictions of two models are quite similar and they are in reasonable agreement with the 
experimental results in the forward rapidity range except for Ar+C interactions, where the models 
underestimate the data.

Deuterons and tritons are produced mostly in the beam fragmentation region for Ar+C and Ar+Al interactions, 
whereas they are mostly produced at mid-rapidity on heavier targets. 
For deuterons, the DCM-SMM and PHQMD models reasonably describe the shape of the experimental spectra 
but underestimate the absolute yields by factors of about 3 and 2, respectively. The triton yields predicted 
by the models are below the experimental data by a factor of about 5.

The $dN/dy$ distributions of protons, deuterons and tritons produced in Ar+A collisions with 
centrality 40--80\% are shown in figures~\ref{yields_yp}(b), \ref{yields_yd}(b) and~\ref{yields_yt}(b). 
The largest contribution is observed in the beam fragmentation region for all the targets.
 This tendency is described by the DCM-SMM and PHQMD models. Again, the models underestimate 
 the absolute yields for deuterons by factors of 3 and 2, respectively. The triton yields predicted 
 by the models are below the experimental data by a factor of about 5.
 A significant deficit of deuterons and tritons in the PHQMD model relative to the experimental data 
 has also been observed in central (0--10\%) Au+Au collisions at $\sqrt{s_{NN}}=3$~GeV by the STAR
 experiment~\cite{STAR_PHQMD}.

The observed discrepancy between the data and the DCM-SMM and PHQMD models could be partially explained by
feed-down from excited nuclear states, which are not taken into account in the models. At BM$@$N 
collision energies, the reaction zone consists of a hadronic gas dominated by 
nucleons and stable nuclei, in particular, $d$, $t$, $^3$He, $^4$He.  However, in addition to these nuclei, 
there are many excited nuclear states with the mass number $A$\,$\geqq$\,4. The role of
the feed-down from these states for the description of light nuclei production in a
broad energy range was discussed in ref.~\cite{feed_vovchenko}. As reported
in~\cite{feed_vovchenko}, feeding gives a significant contribution to the
yields of $d, t$ at NICA/BM$@$N energies: as much as 60\% of all final tritons
and 20\% of deuterons may come from the decays of excited nuclear states. 

\begin{figure}[htpb]
\begin{center}
\vspace{-1.7cm}
\hspace{-0.1cm} (a)

\includegraphics[width=0.64\textwidth,bb=0 0 748 475]{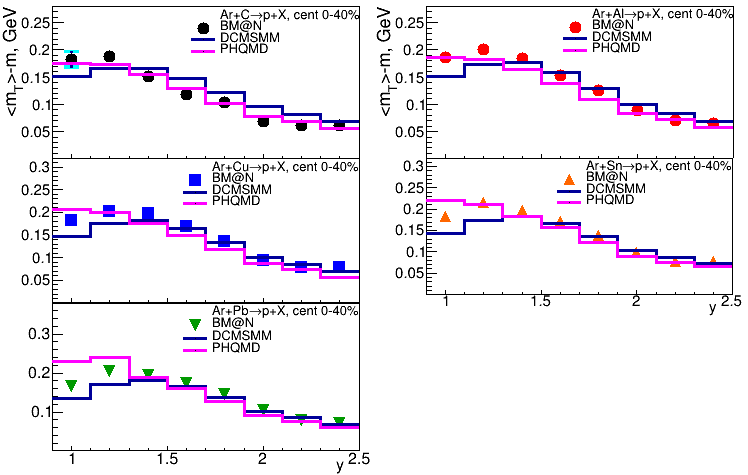}

\hspace{-0.1cm} (b)

\includegraphics[width=0.64\textwidth,bb=0 0 748 475]{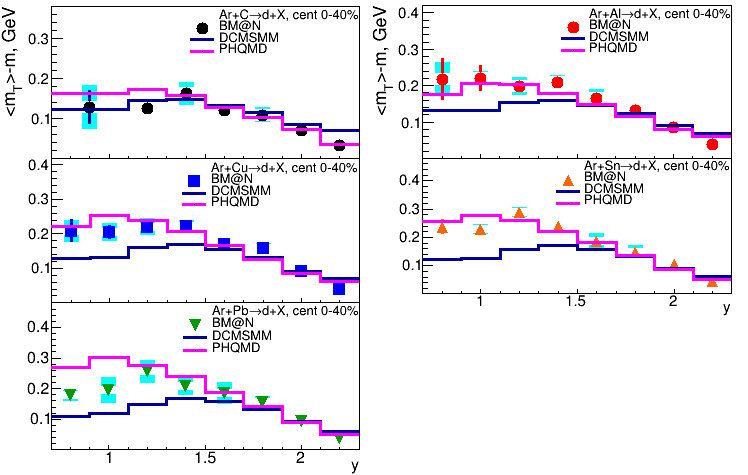}

\hspace{3cm} (c)
\hspace{3cm}

\includegraphics[width=0.64\textwidth,bb=0 0 748 475]{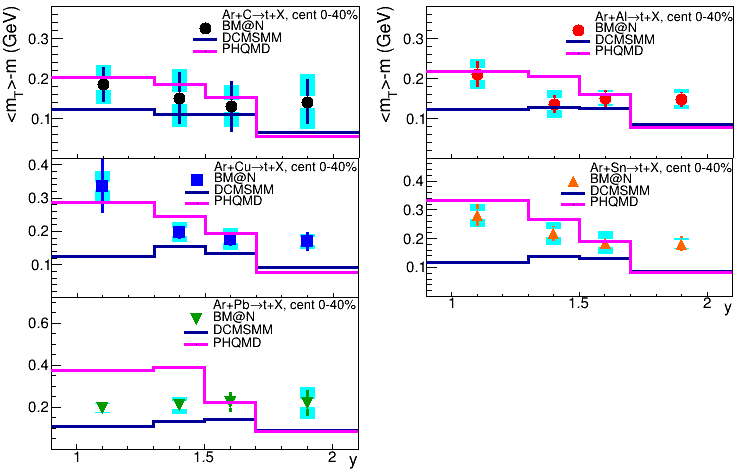}
\end{center}

\vspace{-0.9cm}
 \caption{Mean transverse kinetic energy $\langle E_T\rangle = \langle m_T\rangle - m$ of protons (a), 
 deuterons (b) and tritons (c) in Ar+C, Al, Cu, Sn and Pb interactions with centrality  0--40\% 
 as functions of rapidity $y$.  The vertical bars and boxes represent the statistical and systematic 
 uncertainties, respectively. The predictions of the DCM-SMM and PHQMD models are shown as blue and 
 magenta histograms.}
\label{mtpdt_rapidity}
\end{figure}

The mean transverse kinetic energy, defined as $\langle E_T\rangle = \langle m_T\rangle - m$, 
is related to the $T_0$ value extracted from the fit of the $m_T$ spectrum by the following equation:
\begin{equation}\tag{3}
\langle E_T\rangle = \langle m_T\rangle - m = T_0 + T_0^2/(T_0+m).
\label{Etmean}
\end{equation}
\noindent The $\langle E_T\rangle$ values of protons
in the 0--40\% centrality class are shown in figure~\ref{mtpdt_rapidity}(a) as a function of rapidity.
The maximal values of $\langle E_T\rangle$ are measured at rapidity $1.0<y<1.3$, i.e., at mid-rapidity 
in the CM system. In general, the $y$-dependence of $\langle E_T\rangle$ for protons  is consistent
with  the predictions of the DCM-SMM and PHQMD models.

The $\langle E_T\rangle$ values for deuterons and tritons in the 0--40\% centrality class are shown 
as functions of rapidity in
figures~\ref{mtpdt_rapidity}(b) and \ref{mtpdt_rapidity}(c), respectively.  PHQMD reproduces the 
rise of the data at mid-rapidity in CM for deuterons and tritons
relative to protons, whereas the DCM-SMM model predicts similar  $\langle E_T\rangle$ values f
or protons, deuterons and tritons contrary to the experimental results.

\begin{figure}[htbp]
\begin{center}
 \vspace{-0.1cm}
\includegraphics[width=0.85\textwidth]{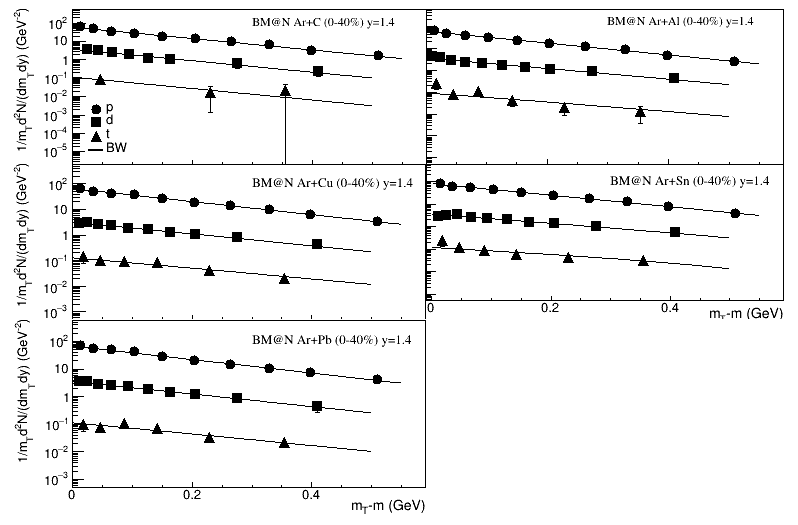}
 \end{center}
\vspace{-0.8cm}
\caption{Invariant $m_T$-spectra of $p,d,t$ produced at rapidity $y$\,=\,1.4 in 0--40\% central Ar+A 
         interactions. The BM$@$N data are shown by various symbols, the fits motivated by the Blast-Wave 
	 model are drawn by lines.}
\label{bw_mt}
\end{figure}

A Blast-Wave model \cite{BWmodel} was used to fit the invariant transverse mass spectra of protons, 
deuterons and tritons according to a formula valid on the assumption of a box-like density profile 
with a uniform density inside the fireball (thermal source) region of transverse radius $r\leq R$:
\noindent
\begin{equation}\tag{4}
 \frac{d^2N}{m_Tdm_Tdy} = \it{Norm(y)}\int_{0}^{R}
         m_TK_1\Big(\frac{m_T\cosh{\rho(r)}}{T}\Big)I_0\Big(\frac{p_T\sinh{\rho(r)}}{T}\Big)rdr, 
\label{BWfunction}
\end{equation}
where $\it {Norm(y)}$ is the normalization factor,  $I_0$ and $K_1$ are the modified Bessel
functions, $T$ is the kinetic freeze-out temperature and $\rho(r) = \tanh^{-1}{\beta (r)}$
is the transverse radial flow rapidity profile. The transverse radial flow velocity $\beta(r)$ 
inside the fireball region is usually parametrized as
$\beta = \beta_s(r/R)^n$, where $\beta_s$ is the fireball-surface velocity. 
Assuming a linear velocity profile (exponent $n=1$), one gets an average transverse radial flow velocity
$\langle\beta\rangle = (2/3)\beta_s$.  The invariant $m_T$-spectra of $p,d,t$  produced at $y=1.4$ in
Ar+C, Al, Cu, Sn and Pb interactions with centrality 0--40\% are shown in figure~\ref{bw_mt}. 
The average radial flow velocity $\langle\beta\rangle$ and source temperature $T$ at the kinetic
freeze-out extracted from the Blast-Wave model fits to the transverse mass spectra of protons, 
deuterons and tritons measured  in the range  $0.9<y<1.5$ ($-0.18<y^*<0.42$ in the center-of-mass system) 
are given in table \ref{table_Tbeta}. The quadratic sum of the statistical and systematical  
uncertainties of data points are used to evaluate the errors of the fit parameters. The parameters 
of the fit were assumed to be constant in the rapidity range of the fit. If a functional form of the 
Boltzmann approximation $T(0)/\cosh{y^*}$ with the midrapidity temperature $T(0)$ is used instead, 
the difference in the fit result is within 5\%. 

One may also obtain the temperature $T$ and mean transverse radial flow velocity 
$\langle\beta\rangle = 2/(n+2)\beta_s$ from common fits of transverse kinetic energies $\langle E_T\rangle$
of protons, deuterons and tritons using the formula derived from eq.~(\ref{BWfunction}) 
in the limit of small $1/z=T/m$ and $\beta_s^2$:

\begin{footnotesize}
\noindent
\begin{align}
\lefteqn{\langle E_T\rangle  = T\Big([1+3/(2z)-9/(8z^2)] + \beta_s^2z[(1+1/z)(1+3/z)-9/(2z^3)]/[2(n+1)]+{}}
                                                                                 \nonumber\\
                & & {}+\beta_s^4z[(3+n(6+5n))+(9+n(18+17n))/z+3(3+n(6+7n))/(8z^2)-{}
                                                                                \nonumber\\
                & & -9(1+n(2+9n))/(8z^3)]/[8(1+n)^2(1+2n)]\Big),   \tag{5}
\label{Etfunction}
\end{align}
\end{footnotesize}
\noindent valid up to terms $O(1/z^3)$ and $O(\beta_s^6)$. Note that at temperatures $T$ of a hundred MeV, 
the $\beta_s^2$-term in eq.~(\ref{Etfunction}) is nearly linear in the cluster mass $m$ down to the proton mass.
The fitted parameters agree with those in table~\ref{table_Tbeta}, except for approximately 50\% larger errors
due to the integration in $\langle E_T\rangle$ of part of the information contained in the $m_T$ spectra.

\begin{figure}[tbp]
\begin{center}
\vspace{-0.5cm}
\includegraphics[width=0.78\textwidth]{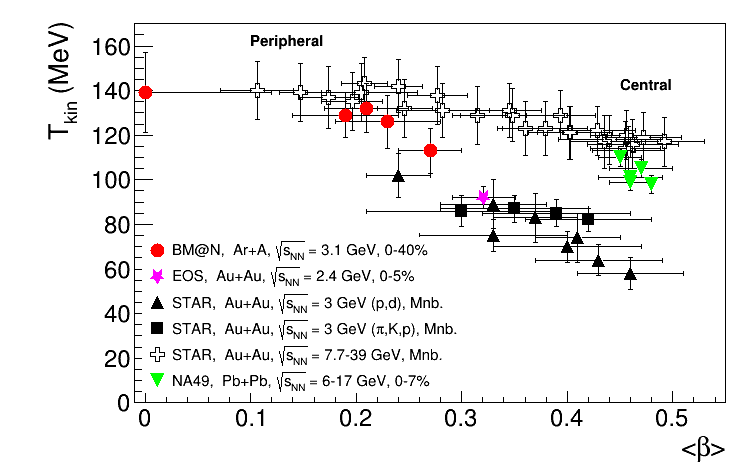}
 \end{center}
\vspace{-0.5cm}
\caption{Kinetic freeze-out parameters ($T_{kin}$ and $\langle\beta\rangle$) in centrality 
         selected nucleus-nucleus collisions: Ar+A (this study); Au+Au from  
EOS~\cite{EOS_radflow} and STAR~\cite{STAR_T0beta,STAR_radflow, star_bw_3gev}; 
Pb+Pb from NA49~\cite{NA49_dtHe3}. For the STAR results, ``Mnb.'' stands for 
minimum bias, the labels ``Peripheral'' and ``Central'' indicate the most peripheral 
(70--80\% central) and the most central (0--5\% central) bins of Au+Au collisions, respectively.}
\label{fig:tkin_beta}
\end{figure}

One finds a flow velocity consistent with zero in central Ar+C collisions. 
Nuclear collisions of such small systems can be considered as a superposition of
independent nucleon-nucleon interactions; therefore, the density of participants reached
in these reactions is probably not high enough to create a fireball with a strong
collective behavior. In contrast, for larger colliding systems (Ar+Al, Cu, Sn and Pb), the
particle density and re-scattering rate inside the reaction zone are higher, giving
rise to a collective flow velocity. It appears that the observed target mass dependence 
for $T$ and  $\langle\beta\rangle$ is weak at BM$@$N energies: fitted temperature and mean 
flow velocity are practically the same within the errors for studied colliding systems. 
This might be an indication 
that the increase of the reaction volume and the number of collisions with the target
mass is not accompanied by a significant compression of the nuclear matter.

The BM$@$N results for kinetic freeze-out parameters ($T_{kin}$ and $\langle\beta\rangle$) could
be compared with measurements at lower and higher energies. Figure~\ref{fig:tkin_beta} presents
results for centrality-selected nucleus-nucleus collisions from the BM$@$N experiment
(this study, 0--40\% central Ar+A at $\sqrt{s_{NN}}$\,=\,3.1 GeV),
the EOS experiment~\cite{EOS_radflow} (0--5\% central Au+Au at $\sqrt{s_{NN}}$\,=\,2.4~GeV), 
the STAR experiment~\cite{STAR_T0beta,STAR_radflow} (0--5\%, 5--10\%, 10--20\%,...,70--80\%  central Au+Au
at $\sqrt{s_{NN}}$\,=\,7.7--39 GeV), and the NA49 experiment~\cite{NA49_dtHe3} (0--7\% central Pb+Pb
at $\sqrt{s_{NN}}$\,=\,6.2--17.3~GeV). Preliminary STAR results
from a Blast-Wave analysis of hadron and light nuclei spectra in centrality-selected Au+Au 
collisions at  $\sqrt{s_{NN}}$ = 3 GeV~\cite{star_bw_3gev} are also presented. These results are shown for different 
combinations of particle species used in the Blast-Wave fits: light hadrons ($\pi, K, p$) or protons
and deuterons ($p,d$). Though the quoted uncertainties in a Blast-Wave motivated analysis are large, 
there is an indication that the system size trend for kinetic freeze-out parameters is different 
in low ($\sqrt{s_{NN}}<6$~GeV) and high-energy collisions.
 
\begin{table}[!hbp]
\vspace{-0.5cm}
  \caption {
$T$ and $\langle\beta\rangle$ values evaluated from the Blast-Wave fit of the transverse mass spectra 
of protons, deuterons and tritons produced in the CM system rapidity range $-0.18<y^*<0.42$ 
in Ar+A interactions with centrality 0--40\%. The errors represent the uncertainties of the fit 
to the data points with the quadratic sum of the statistical and systematical uncertainties.}
\begin{footnotesize}    
\vspace{0.3cm}
\begin{tabular}{|l|c|c|c|c|c|}
\hline
& & & & & \\
& Ar+C & Ar+Al & Ar+Cu & Ar+Sn & Ar+Pb \\
& & & & & \\
\hline
& & & & & \\
T , MeV                   &  $140\pm18$ & $129\pm10$ & $132\pm11$ & $113\pm10$  & $126\pm12$  \\
& & & & & \\
$\mathbf \langle\beta\rangle$   &  $0.0\pm^{0.12}_{0.0}$  & $0.19\pm0.05$ & $0.21\pm0.04$ & $0.27\pm0.03$ & $0.23\pm0.05$ \\
& & & & & \\
$\mathbf \chi^2/ndf$            &  44/49  &  127/55       &   113/55      &   86/55       & 172/55 \\
& & & & & \\
\hline
\end{tabular}
\end{footnotesize}
\label{table_Tbeta}
\end{table}

\section{Coalescence factors}
\label{sect66}

Within a coalescence model~\cite{Coal1,Coal2,Sat81}, nuclear fragment  formation is characterized 
by a coalescence factor $B_A$, defined
through the invariant momentum spectra by the equation:

\begin{small}
\hspace{1.5cm}
$E_A d^3 N_A / d^3 p_A = B_A  (E_p d^3 N_p / d^3 p)^Z  (E_n d^3 N_n / d^3 p)^{A-Z}_{\vert p=p_A/A}$ , 
\end{small}

\noindent where $p_A$ and $p=p_A/A$ are momenta of the nuclear fragment A and the nucleon, respectively. 
It relates the yield $N_A$ of nuclear fragments with charge $Z$ and atomic mass number $A$ to the yields
 of the coalescing nucleons $N_p$ and $N_n$ at the same velocity.  Assuming that
neutron momentum density is equal to the proton
momentum density at freeze-out, the $B_A$ value can be calculated as:
\begin{small}
\begin{equation}\tag{6}
  B_A = d^2 N_A / 2\pi p_{T,A}dp_{T,A}dy / (d^2 N_p / 2\pi p_T d p_T dy)^A / (n/p)^{A-Z} , 
\end{equation}
\end{small}

\noindent where $n/p$ is the ratio of the numbers of produced neutrons to protons, 
$p_{T,A}$ and $p_T$ are the transverse momenta of the nuclear fragment A and the proton, respectively. 
The coalescence factor is inversely related to the effective emission volume of the
nucleons with nearby 3-momenta
\cite{Sat81}: $B_A \sim V_{eff}^{ 1-A}$. The strong position-momentum
correlations present in the expanding source lead to a higher coalescence probability at larger $p_T$ values. 
Assuming a box-like transverse
density profile of the source, the model predicts at small or moderate $p_T$ \cite{Scheibl}:
\begin{small}
\begin{equation}\tag{7}
\label{eq7}
B_A \simeq g_s \Lambda_A A^{-1/2}C_A[(2\pi)^{3/2}/(m_T R_{\parallel}(m_T)R_{\bot}^2(m_T))]^{A-1}\exp[m_T (1/T_p -1/T_A)],     
\end{equation}
\end{small}

\noindent where $g_S=(2S+1)/2^A$ is the spin factor of the nuclear fragment A, $\Lambda_A$ is a 
suppression factor of correlated nucleons, e.g., due to a feed-down fraction of uncorrelated nucleons 
produced in hyperon decays, $C_A$ is a quantum correction factor related to the finite fragment 
size~\cite{Sat81,Scheibl}, $R_{\bot}$ and $R_{\parallel}$ are the femtoscopic radii of the source 
in the longitudinally co-moving system  (LCMS)~\cite{Scheibl}, $T_p$ and $T_A$ are the inverse 
transverse momentum slopes for proton and fragment A, respectively. 
The $\Lambda_A$ factor is close to 1 in the BM$@$N energy range, as the fraction of nucleons 
originated from hyperon decays is around 2\% according to predictions of the UrQMD model~\cite{UrQMD}. 
The UrQMD and PHQMD models predict the $n/p$ ratio to be between 1.09 and 1.18 in the BM$@$N rapidity 
range for Ar+C and Ar+Pb interactions, respectively (see also section \ref{section_stopping}).  
\begin{figure}[htpb]
\begin{center}
\vspace{-0.5cm}
\hspace{-0.5cm} (a)

\hspace{-0.5cm}
\includegraphics[width=0.9\textwidth,bb=0 0 748 475]{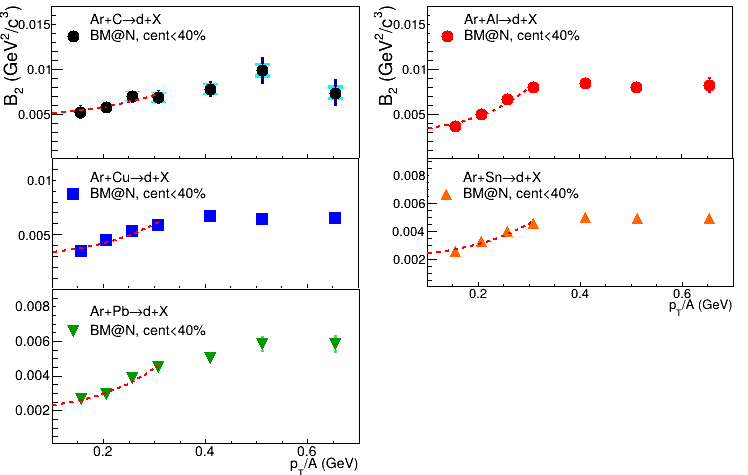}

\hspace{-0.5cm} (b)
\hspace{-0.5cm}

\includegraphics[width=0.85\textwidth,bb=0 0 748 475] {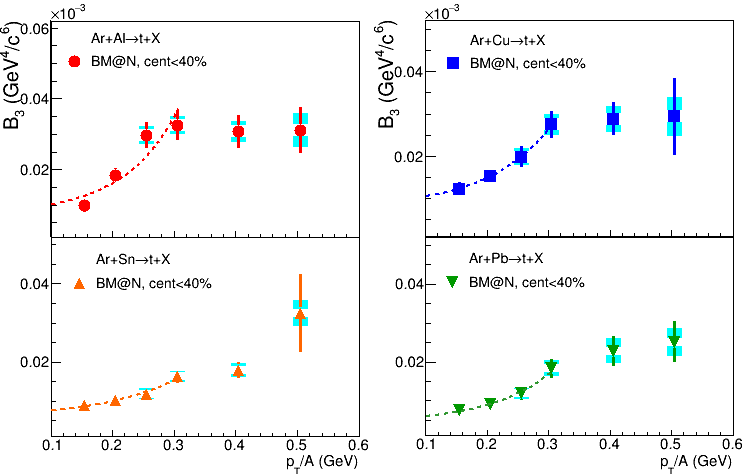}
\vspace{-0.5cm}
\end{center}
 \caption{Coalescence parameter $B_2$ for deuterons (a) and  $B_3$ for tritons (b) measured as a 
 function of $p_T/A$ in the rapidity range $-0.18<y^*<0.62$ in Ar+A collisions
 with centrality 0--40\%. Dash lines show results of the fits in the range $p_T/A < 0.32$ described in the text.}
 \label{B23pT}
\end{figure}

The $B_2$ and $B_3$ values as functions
of the transverse momentum measured in  argon-nucleus interactions with centrality
0--40\% are shown in figures~\ref{B23pT}(a) and $\ref{B23pT}$(b), respectively. The transverse 
momentum is scaled to the atomic number of the nuclear fragment
(deuteron, triton), $p_T/A$. The yields of protons ($N_p$), deuterons ($N_d$) and
tritons ($N_t$)  are measured in the same rapidity range, namely
$0.9<y<1.7\, (-0.18<y^*<0.62)$. The statistics of tritons
is not sufficient to present $B_3$ for Ar+C interactions.  It is found that $B_2$ and $B_3$ 
rise with $p_T$ at low $p_T$ and saturate at higher $p_T$ for all the targets used in measurements.
The $B_2$ and $B_3$ values at low $p_T$ are smaller for heavier targets compared to lighter targets. 

\begin{table}[!tbp]
\vspace{-0.9cm}

  \caption {
Coalescence parameters  $B_2(p_T=0)$ and $B_3(p_T=0)$  extrapolated to $p_T=0$ using an exponential fit to $B_2(p_T)$
and $B_3(p_T)$; coalescence radii $R^d_{coal}(p_T=0)$ and $R^t_{coal}(p_T=0)$  evaluated
from the $B_2(p_T=0)$ and $B_3(p_T=0)$ values for deuterons
and tritons produced in the rapidity ranges $-0.18<y^*<0.22$ and $0.22<y^*<0.62$ in Ar+A interactions 
with centrality 0--40\%. The quoted errors are the quadratic sums of the statistical and systematic uncertainties.}
\begin{footnotesize}
\vspace{0.3cm}
\begin{tabular}{|l|c|c|c|c|c|}
\hline
& & & & & \\
& Ar+C & Ar+Al & Ar+Cu & Ar+Sn & Ar+Pb \\
& & & & & \\
\hline
                                             $-0.18<y^*<0.22$
 & & & & & \\
 & & & & & \\
$B_2(p_T=0)/10^3$, GeV$^2$/c$^3$    & $3.2\pm1.0$ & $1.95\pm0.7$ & $2.6\pm0.3$ & $1.8\pm0.2$  & $1.35\pm0.2$  \\
& & & & & \\
$B_3(p_T=0)/10^6$, GeV$^4$/c$^6$    &  --           & $7.2\pm2.2$ & $5.8\pm2.8$ & $4.9\pm0.6$  & $2.6\pm0.4$ \\
& & & & & \\
$R^d_{coal}(p_T=0)$, fm    & $2.3\pm0.3$ & $2.7\pm0.3$ & $2.5\pm0.2$ & $2.8\pm0.2$  & $3.1\pm0.2$  \\
& & & & & \\
$R^t_{coal}(p_T=0)$, fm    &    --      & $2.4\pm0.2$ & $2.5\pm0.2$ & $2.5\pm0.2$  & $2.9\pm0.2$ \\
& & & & & \\
\hline
                                            $0.22<y^*<0.62$
& & & & & \\
& & & & & \\
$B_2(p_T=0)/10^3$, GeV$^2$/c$^3$    & $4.07\pm1.0$ & $3.56\pm0.5$ & $3.0\pm0.8$ & $2.06\pm0.5$  & $2.67\pm0.4$  \\
& & & & & \\
$B_3(p_T=0)/10^6$, GeV$^4$/c$^6$    &   --          & $9.6\pm3.0$ & $9.3\pm2.9$ & $7.3\pm2.7$  & $5.1\pm2.3$ \\
& & & & & \\
$R^d_{coal}(p_T=0)$, fm    & $2.1\pm0.2$ & $2.2\pm0.2$ & $2.4\pm0.2$ & $2.7\pm0.2$  & $2.5\pm0.2$  \\
& & & & & \\
$R^t_{coal}(p_T=0)$, fm    &     --        & $2.2\pm0.2$ & $2.3\pm0.2$ & $2.4\pm0.2$  & $2.5\pm0.2$ \\
& & & & & \\
\hline
\end{tabular}
\end{footnotesize}
\label{table_B2B3}
\end{table}

\begin{figure}[htbp]
\begin{center}
 \vspace{-0.5cm}
\hspace{-2cm} (a)
\hspace{-2cm}

\includegraphics[width=0.9\textwidth,bb=0 0 1382 653]{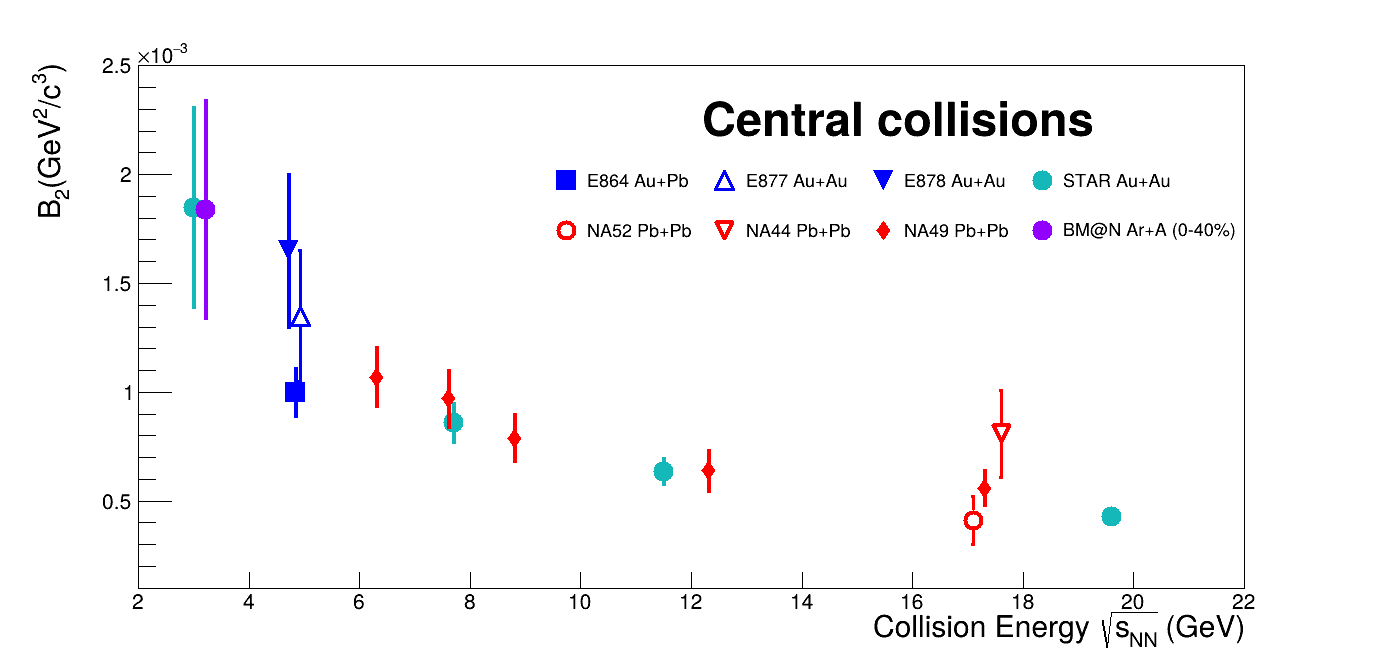}

\hspace{0cm} (b)
\hspace{0cm}

\includegraphics[width=0.9\textwidth,bb=0 0 1382 653]{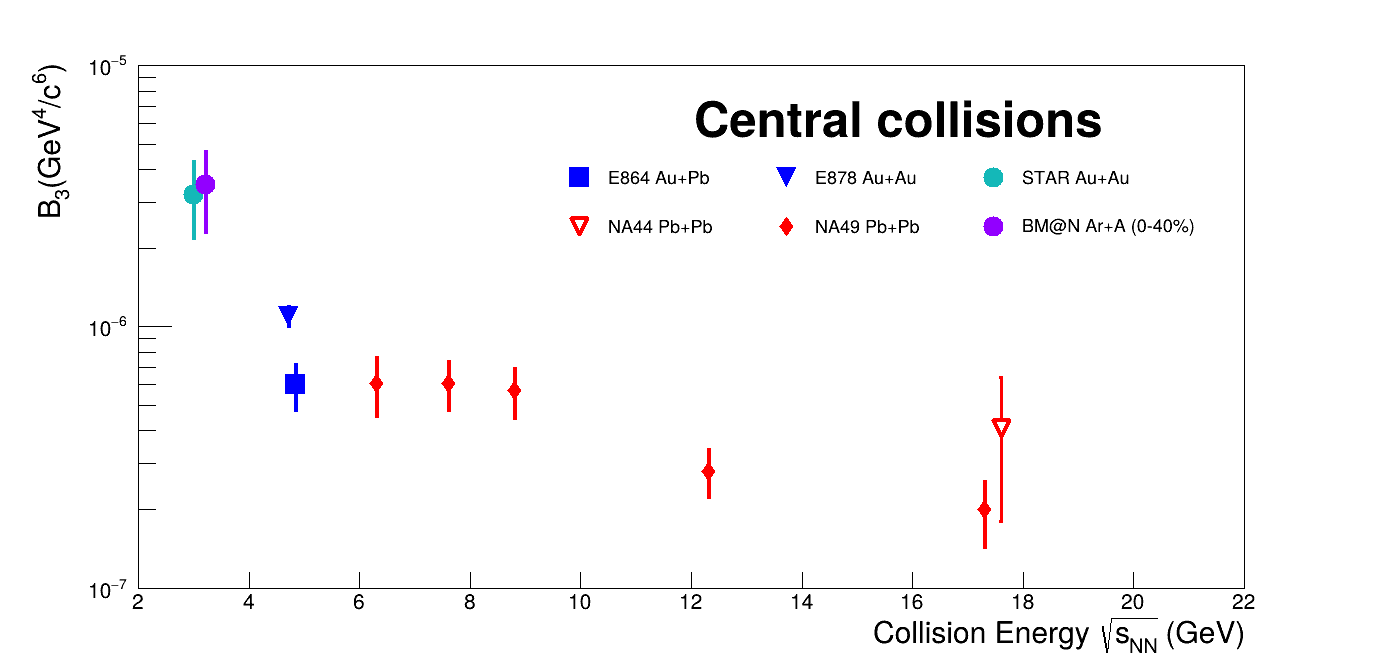}
 \end{center}
\vspace{-0.5cm}
\caption{Coalescence parameters $B_2(p_T=0)$ (a) and $B_3(p_T=0)$ (b) for deuterons and 
tritons as functions of the nucleon-nucleon center-of-mass energy. The BM$@$N result is the 
weighted average value calculated in the rapidity range $-0.18<y^*<0.22$  for Ar+Al, Cu, Sn and Pb 
interactions with centrality 0--40\%.}
\label{B2B3_compare}
\end{figure}

In order to compare the present measurements of $B_2$ and $B_3$ with previously obtained results,
the $B_2(p_T)$ and $B_3(p_T)$ values given in figure~\ref{B23pT}  are 
extrapolated down to $p_T = 0$ using exponential fits of the form $b\exp[a(m_T-m_A)]$ as predicted 
by the coalescence model with a box-like density profile \cite{Scheibl}, see eq.~(\ref{eq7}). 
The fits are performed for the first four data points in the range $p_T/A < 0.32$. 
If the fit results with $\chi^2/ndf>1$, the uncertainty of the parameter $B_A(p_T = 0)$ 
is scaled up by a factor $\sqrt{\chi^2/ndf}$ following a recommendation given in ref.~\cite{PDG2010}. 
The results of the extrapolation are presented in table~\ref{table_B2B3}.

The BM$@$N  values of $B_2= 1.84\pm 0.5$ GeV$^2$/c$^3$ and $B_3= 3.5\pm 1.2$ GeV$^4$/c$^6$ calculated 
as the weighed average values for Ar+Al, Cu, Sn and Pb interactions with centrality 0--40\% are compared in
 figure~\ref{B2B3_compare} with the results of other experiments: 
 STAR (0--10\% central, $p_T/A=0.65~{\rm GeV/c}$)~\cite{STAR_PHQMD,STAR_pdt,STAR_LightNucl},
  NA44 (0--10\% central) \cite{Bearden}, NA52 \cite{NA52}, E864 \cite{E864}, E877 \cite{E877}, 
  E878 \cite{E878} (0-10\% cental), NA49 (0--7\% central) \cite{NA49_dtHe3}.  
The $B_2$ and $B_3$  results for Ar+A interactions with centrality 0--40\% are consistent 
with the general trend of decreasing $B_2$ and $B_3$ values with increasing collision energy of 
central interactions of heavy nuclei.
The $B_2$ and $B_3$ values are inversely related to the coalescence radius $R_{coal}$, which 
is closely related to the LCMS femtoscopic radii of the source $R_{out}, R_{side}, R_{long} = R_{\parallel}$
 with $R_{out}(p_T=0) = R_{side}(p_T=0) = R_{\bot}$~\cite{Scheibl}.
On the basis of eq.~(\ref{eq7}) at $p_T = 0$, one can define
$R_{coal}=\sqrt[3]{R_{\parallel}R_{\bot}^2}$  and calculate it from the $B_2(p_T=0)$ and $B_3(p_T=0)$ 
values of deuterons and tritons. In the calculations, the $C_d$ and
$C_t$ factors from \cite{Bearden} are scaled according to the mass of the colliding systems to 
account for the suppression related to the increased effective volume due to the finite deuteron and 
triton radii, see eq.~(4.12) in ref.~\cite{Scheibl}. The resulting values are in the range of 0.55--0.61 
and 0.51--0.58 for $C_d$ and $C_t$,
respectively. The results for $R_{coal}$ are also given in table~\ref{table_B2B3}. 
\begin{figure}[htbp]
\begin{center}

\includegraphics[width=0.9\textwidth,bb=0 0 1382 653]{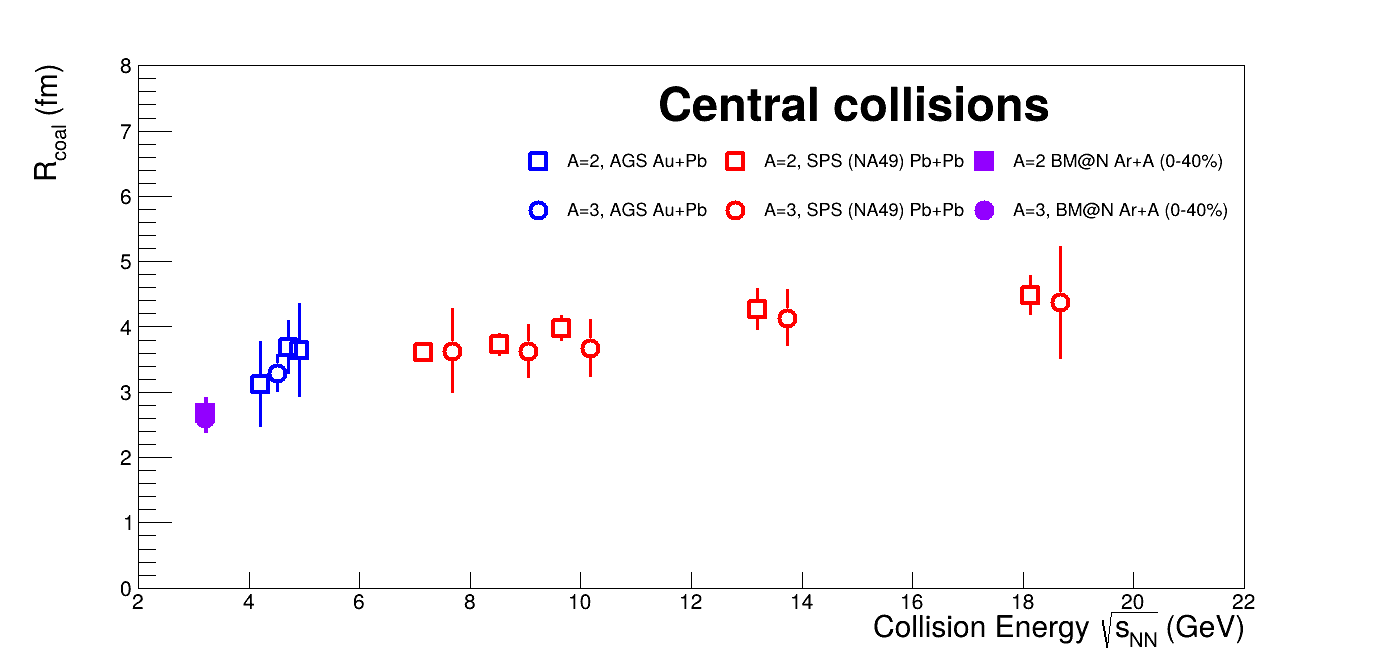}
 \end{center}
\vspace{-0.5cm}
\caption{Coalescence radii $R_{coal}$ for deuterons and tritons as a function of the
nucleon-nucleon center-of-mass energy. The BM$@$N result is the weighted average value
calculated in the rapidity range $-0.18<y^*<0.22$ for Ar+Al, Cu, Sn and Pb interactions with centrality 0--40\%.}
\label{Rad_compare}
\end{figure}

The coalescence source radii for deuterons and tritons produced in Ar+Al, Cu, Sn and Pb  interactions 
with centrality 0--40\% are consistent within the uncertainties. 
The BM$@$N values for the coalescence radii averaged for Ar+Al, Cu, Sn and Pb interactions are compared 
in figure~\ref{Rad_compare} with results obtained at higher energies and for larger collision 
systems as compiled in ref.~\cite{NA49_dtHe3}.
A weak increase of the coalescence radii as a function of the center-of-mass energy in 
the nucleon-nucleon system is seen in figure~\ref{Rad_compare}. 
One can conclude that the BM$@$N results reported here are consistent with no or weak dependence 
of $R_{coal}$ on the target size within the measurement uncertainties.

\section{Baryon rapidity distributions, stopping and rapidity loss in Ar+A}
\label{section_stopping}

\hspace{6mm} The total baryon number at a given rapidity in Ar+A collisions at NICA/\allowbreak BM$@$N energies
is basically determined by the nucleons and the light nuclei ($d, t, ^3$He). According to the results on 
the rapidity spectra of protons and light nuclei presented in section~\ref{sect6}, the
number of nucleons bound in clusters contributes to the total number of baryons up to about
15\% and 25\% in central Ar+C and Ar+Pb reactions, respectively. To obtain the baryon 
rapidity distribution, we add up the baryon number of the measured protons, deuterons and 
tritons in each rapidity bin. The obtained distribution is then corrected for unmeasured 
baryons: neutrons, hyperons and $^3$He nuclei. Calculations with the PHQMD and UrQMD models 
indicate that for all collision systems, the $n/p$~ratio is about 1.1 in the forward 
hemisphere, varying slowly with rapidity and then increasing abruptly to $\approx$1.22 (the
$n/p$~ratio in the projectile Ar~nucleus) at the beam rapidity. We use these model 
predictions to estimate the yield of neutrons $n$; furthermore, we assume that the $t/^3$He 
ratio is equal to $n/p$. Hyperons contribute less than 2\% to the total baryon number according 
to the PHQMD and UrQMD~\cite{UrQMD} models and 
are thus neglected. The total number of baryons $B$ in a rapidity bin is then calculated as
\begin{center}
$B = p + n + 2\cdot d + 5.7\cdot t$, 
\end{center}
where the coefficient in front of $t$ is 5.7. It is calculated as the sum of 3 for tritons and 3/1.1 for $^3$He.
\begin{figure}[tpb]
 \begin{minipage}[h]{0.5\linewidth}
    \includegraphics[width=70mm,angle=0]{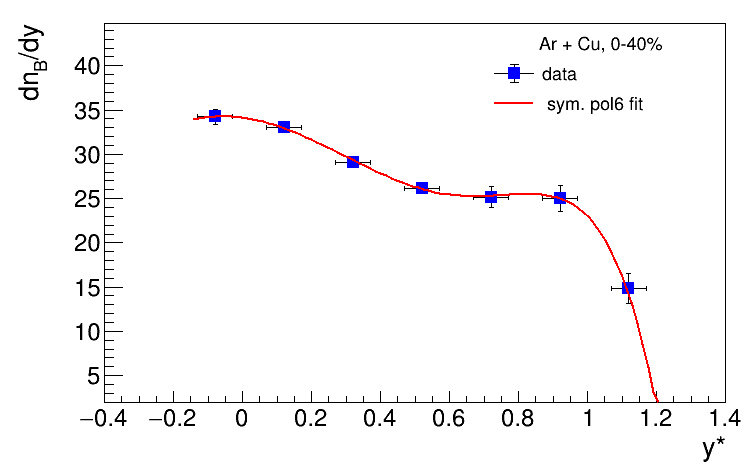}
 \end{minipage}
 \begin{minipage}[h]{0.5\linewidth}
   \includegraphics[width=70mm,angle=0]{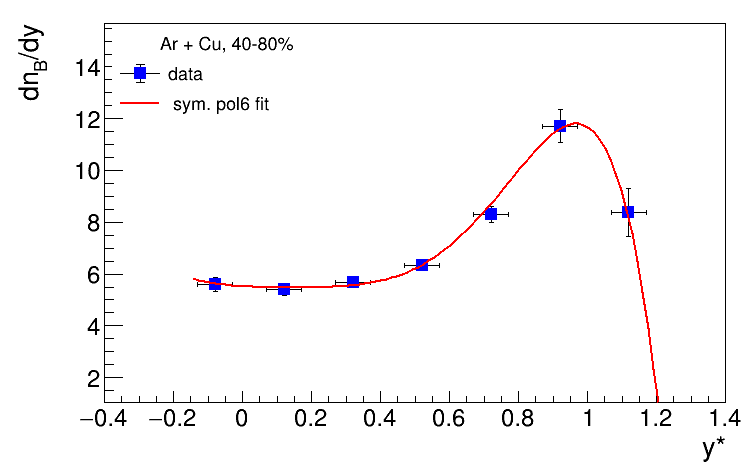}
 \end{minipage}
 \caption{Left: Rapidity distribution of baryons in 0--40\% central Ar+Cu collisions. 
 The measurements are shown by points, whereas the solid line represents the results of a 
 fit with a 3$^{rd}$ order polynomial in $y^{*2}$. Right: same, but for 40--80\% central Ar+Cu collisions.}
 \label{fig5}
\end{figure}
The resulting baryon rapidity distributions for Ar+Cu collisions are shown in 
 figure~\ref{fig5} as a function of the rapidity in the center-of-mass system $y^{*}$: 
 the left panel shows the results for 0--40\% central collisions, and the right one is 
 for 40--80\% central collisions.
 A large difference in the shapes of the d$n$/d$y$ distributions is observed as more baryons 
 are transported to midrapidity in the more central collisions. 
 In order to describe
 those shapes, the data were fitted  by a 3$^{rd}$ order polynomial in $y^{*2}$, as
 suggested in ref.~\cite{brahms200}. The results of the fit  are shown in figure~\ref{fig5} by solid line curves.

The average rapidity loss is calculated (below $y$\,=\,$y^*$) as:
\begin{equation}\tag{8}
\label{eq3}
\langle \delta y \rangle = y_{b} - \langle y \rangle,
\end{equation}
\noindent where $y_b$\,=\,1.08 is the rapidity of the projectile in the center-of-mass
system, and the average rapidity:
\begin{equation}\tag{9}
\label{eq9}
\langle y \rangle =  \int_{0}^{y_{b}} y \frac{dn}{dy} dy  \Biggm/ \int_{0}^{y_{b}} \frac{dn}{dy}dy\ .
\end{equation}

This equation refers to net-baryons, i.e. baryons minus antibaryons. At NICA energies, 
however, the production of antibaryons is so rare that the difference between baryons and 
net-baryons is negligible.

\begin{figure}[tpb]
\centering
\includegraphics[width=0.8\textwidth]{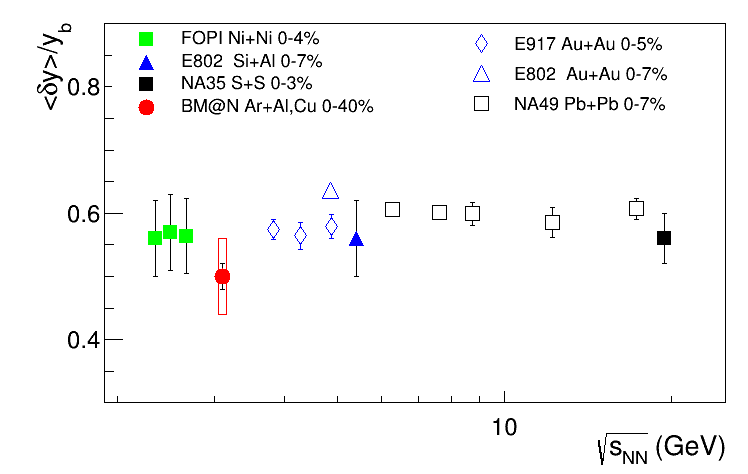}
\caption{\label{deltay_excitation_func}The excitation function of the scaled average 
rapidity loss $\langle \delta y \rangle$/$y_{b}$ in nucleus-nucleus collisions. Medium-size 
colliding systems~\cite{rap_loss_vide, stop_fopi, stop_na35} are drawn by solid symbols, while heavy
systems~\cite{rap_loss_vide, stop_e917,sps_deltay} are shown by open ones. 
Centrality intervals are indicated in the legends. The BM$@$N data point is the 
average of Ar+Al and Ar+Cu results, its systematic error is shown by the box.}
\end{figure}
\begin{table}
\centering
\caption{\label{table3} The average rapidity loss $\langle \delta y \rangle$ in Ar+A reactions. 
          The quoted uncertainties are statistical errors.} 
\begin{tabular}{|c|c|c|c|c|c|}
\hline
& Ar+C & Ar+Al & Ar+Cu & Ar+Sn & Ar+Pb \\
\hline
0-40\% & 0.42 $\pm$ 0.04 &  0.50 $\pm$ 0.03 & 0.58 $\pm$ 0.02 & 0.63 $\pm$ 0.02 & 0.65 $\pm$ 0.02 \\
40-80\% & 0.38 $\pm$ 0.04 &  0.41 $\pm$ 0.04& 0.45 $\pm$ 0.03& 0.47 $\pm$ 0.03& 0.48 $\pm$ 0.04\\
\hline
\end{tabular}
\end{table}

The $\langle \delta y \rangle$ values for 0--40\% central and 40--80\% central Ar+A collisions are
listed in table~\ref{table3}. A clear trend is observed: $\langle \delta y \rangle$
increases with the target mass and with collision centrality. This behavior is expected because the
probability of multiple interactions in the projectile-target overlap region also rises
with the centrality and target mass. The quoted (statistical) uncertainties are the
standard errors of the mean $\langle y \rangle$ calculated from the data points
within the rapidity range [$0,y_b$]. The systematic error of the rapidity loss 
values comes from the uncertainty in the fitting procedure used to describe the baryon 
rapidity spectra.
This uncertainty is taken as the difference between the
total baryon number estimated from the fit function and the one obtained from the data points.
It varies from 7\% to 12\%.

The energy dependence of the scaled average rapidity 
shift $\langle \delta y \rangle$/$y_{b}$ in nucleus-nucleus collisions as a function of 
$\sqrt{s_{NN}}$ is shown in figure~\ref{deltay_excitation_func}. The average of the BM$@$N 
results obtained in Ar+Al and Ar+Cu collisions is shown together with results from medium-size 
almost symmetric colliding systems 
from~\cite{rap_loss_vide, stop_fopi, stop_na35} (solid symbols) and those
from heavy colliding systems~\cite{rap_loss_vide, stop_e917, sps_deltay} (open symbols). 
The corresponding  centrality intervals are indicated in the legends.
As one can see, the scaled rapidity loss does not vary significantly over a broad energy range. 

\section{Particle ratios}
\label{sect67}
\begin{figure}[tpb]
\centering
\includegraphics[width=0.95\textwidth]{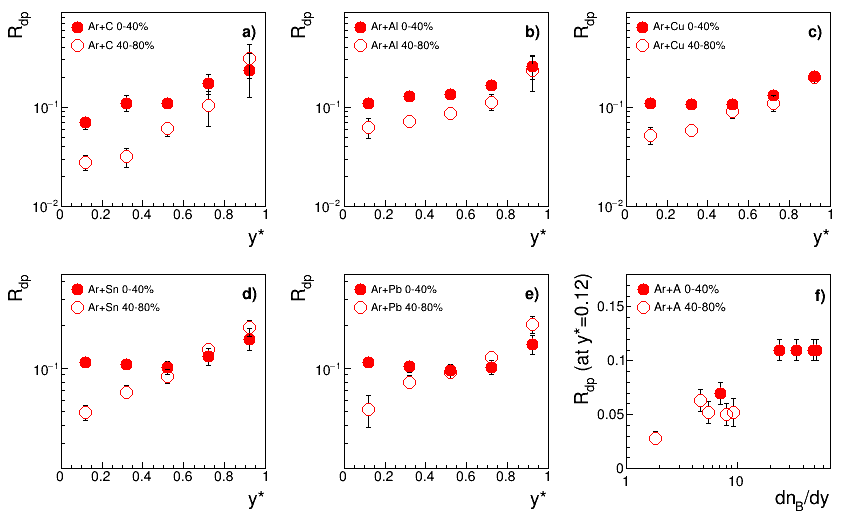}
\caption{\label{dp_rapidity}$R_{dp}$ as a function of center-of-mass rapidity $y^{*}$ in Ar+C 
(a), Ar+Al (b), Ar+Cu (c), Ar+Sn (d), and Ar+Pb (e) collisions. Results for collisions with centrality of 0-40\% and 40-80\% 
 are shown by solid and open symbols, respectively. Panel~(f): midrapidity $R_{dp}$
as a function of midrapidity baryon density d$n_B$/d$y$ in Ar+A collisions.}.
\end{figure}

\hspace{6mm}The rapidity and centrality dependence of the deuteron-to-proton ratio
$R_{dp}$ in Ar+A collisions at 3.2~A~GeV ($\sqrt{s_{NN}}$\,=\,3.1~GeV) is presented in
figure~\ref{dp_rapidity}~(a)--(e). Collisions with centrality 0--40\% central and 40--80\% are represented
by solid and open symbols, respectively.
As one can see, $R_{dp}$ rises strongly from midrapidity to the beam rapidity in more
peripheral collisions. The same trend is observed in 0--40\% central Ar+C collisions. In contrast, 
in 0--40\% central collisions of argon nuclei with aluminum or heavier targets, 
$R_{dp}$ indicates a plateau-like behavior near midrapidity followed by an increase toward 
the beam rapidity region. The plateau region for $R_{dp}$ increases gradually with the 
target mass number covering almost all the measured rapidity range in Ar+Pb collisions.

The midrapidity $R_{dp}$ values from Ar+A collisions with centrality 0--40\%  and 40--80\%  as functions
of the midrapidity baryon density d$n_B$/d$y$ (obtained from the fits of figure~\ref{fig5})
are presented in figure~\ref{dp_rapidity}~(f). As can be seen, $R_{dp}$ increases steadily
for low values of d$n_B$/d$y$ and then levels off at higher values.

For a system in chemical equilibrium and a size
substantially larger than
the deuteron radius, the ratio of the invariant yield of deuterons to the one of protons can be
related to the average proton phase-space density at freeze-out $\langle f_p \rangle$ as
\begin{equation}\tag{10}
\label{eq_phasespace}
\langle f_p \rangle = \frac{R_{pn}}{3}\frac{E_d\frac{d^3N_d}{d^3P}}{E_p\frac{d^3N_p}{d^3p}}\ ,
\end{equation}
where $R_{pn}$ is the proton-to-neutron ratio, $P$\,=\,2$p$, and the factor of 3 accounts for 
the particle spins~\cite{na44_phasespace}. The $\langle f_p \rangle$ value depends
on the strength of the nuclear stopping in the reaction as well as on the outward flow
effects. 

\begin{figure}[htpb]
 \begin{minipage}[h]{0.5\linewidth}
    \includegraphics[width=70mm,angle=0]{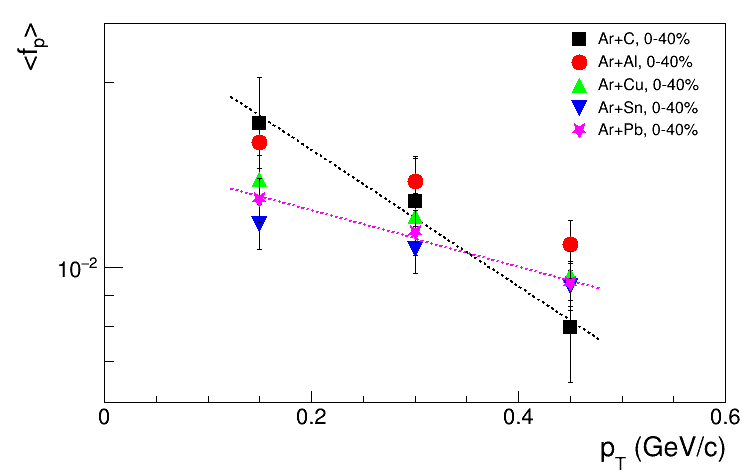}
 \end{minipage}
 \begin{minipage}[h]{0.5\linewidth}
   \includegraphics[width=70mm,angle=0]{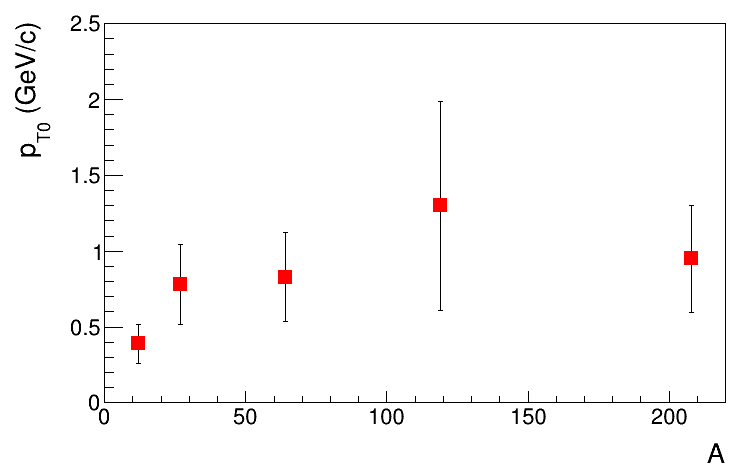}
 \end{minipage}
 \caption{Left: Average proton phase-space density for 0--40\% central Ar+A collisions as a 
 function of $p_T$  within the rapidity range 0.02$<y^*<$0.42. Dashed lines show
 fits to exponent (see text for details). Right: The inverse slope parameter $p_{T0}$ of the
 $p_T$-dependence of $\langle f_p \rangle$ as a function of the target mass number.
 }
 \label{fig_dp}
\end{figure}

The $p_T$-dependence of the average proton phase-space
density is shown in the left panel of figure~\ref{fig_dp}. Here, the ratio of deuterons to protons 
is obtained in the  rapidity range
0.02\,$<$\,$y^*$\,$<$\,0.42 and at three $p_T/A$ values: 0.15, 0.3, and  0.45~GeV/c. 
The $\langle f_p \rangle$ values are calculated according to eq.~(\ref{eq_phasespace}).
The values of the $R_{pn}$ ratio in the chosen phase-space region were taken from the UrQMD model.
As one can see, $\langle f_p \rangle$ decreases with $p_T$ in all 
reaction systems. Such a trend is indeed expected for a thermal source at a low phase-space density
($f_p$\,$<<$\,1), where $\langle f_p \rangle$ follows a Boltzmann distribution and
decreases exponentially with $p_T$~\cite{f_pt}. The dashed lines in figure~\ref{fig_dp} show fits by
the exponential function $const \cdot \exp(-p_T/p_{T0})$ for $\langle f_p \rangle$ from Ar+C
and Ar+Pb reactions with $p_{T0}$ as the inverse slope parameter. It is known that the presence of outward
flow in the system makes $f(p_T)$ flatter as the radial velocity increases~\cite{f_flow}.
The right panel of figure~\ref{fig_dp} shows the system-size dependence of the slope parameter
$p_{T0}$ of the $p_T$-dependence for $\langle f_p \rangle$.
As one can see, this dependence is, indeed, correlated with the results on the radial velocity 
presented in table~\ref{table_Tbeta}: i.e., weak radial expansion in Ar+C and approximately the 
same strength of collective radial flow in Ar+Al, Cu, Sn and Pb.

It was identified long time ago that the nuclear cluster abundances and the entropy value 
attained in the collisions are related. According to an early investigations~\cite{simens_kapusta}, 
in a mixture of nucleons and deuterons in thermal and chemical equilibrium, the entropy per 
nucleon $S_N/A$ can be deduced from the deuteron-to-proton ratio $R_{dp}$ as
\begin{equation}\tag{11}
\label{eq11}
\frac{S_N}{A} = 3.945 - \ln{R_{dp}} - \frac{1.25 R_{dp}}{1 + R_{dp}}.
\end{equation}

Furthermore, as the collision energy increases, the contribution of mesons $S_{\pi}$ to 
the total entropy becomes important. Following~\cite{landau}, the entropy of pions per
nucleon can be estimated by
\begin{equation}\tag{12}
\label{eq12}
\frac{S_{\pi}}{A} = 4.1\frac{N_{\pi}}{N_N},
\end{equation}
where $N_N$\,=\,$N_p$\,+\,$N_n$ is the total number of nucleons.

\begin{figure}[htpb]
\centering
\includegraphics[width=0.9\textwidth]{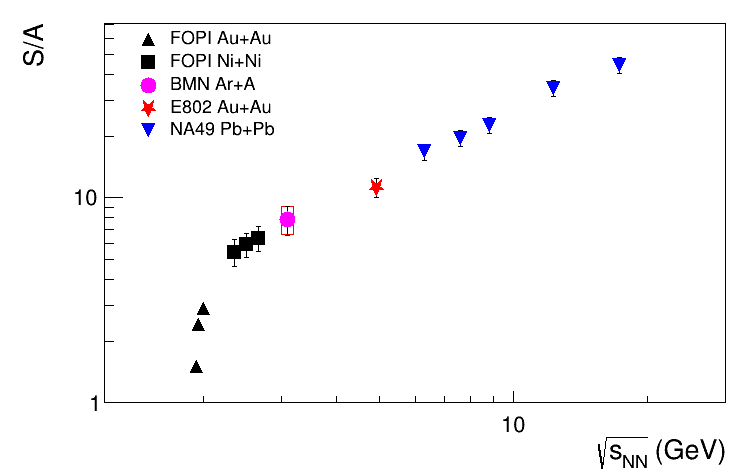}
\caption{\label{entropy_energy}The excitation function of the entropy per baryon $S/A$
from SIS/FOPI\cite{stop_fopi, fopi_au}, AGS/E802\cite{e802_entr}, SPS/NA49\cite{NA49_dtHe3, na49_pika_40, 
na49_pika_20, NA49_deut} and NICA/BM$@$N (this study).}
\end{figure}

We thus calculated the total entropy $S/A$ near midrapidity as the sum of the nucleon and
pion entropy contributions according to eqs.~(\ref{eq11}) and~(\ref{eq12}). To estimate
$S_{\pi}$, we used the recently published BM$@$N results on positively charged 
pions~\cite{BMN_piKpaper}, while the contribution of $\pi^-$, $\pi^0$, and neutrons was 
obtained from the UrQMD model. We found that the contribution of pions to the total entropy
does not exceed 25\% in Ar+A collisions at NICA energies. Finally,  $S/A$ is found to be 
10.3, 7.8, 7.8, 7.9, and 7.9 in central Ar+C, Ar+Al, Ar+Cu, Ar+Sn, and Ar+Pb, respectively.
The estimated uncertainty in $S/A$ is about 15\%. In figure~\ref{entropy_energy} 
the energy dependence of $S/A$ in central heavy-ion collisions is presented. This compilation includes 
data from experiments that have published numerical values for the midrapidity yields of 
charged pions, protons, and light  nuclei~\cite{NA49_dtHe3, stop_fopi, fopi_au, e802_entr, na49_pika_40, 
na49_pika_20, NA49_deut}.
The BM$@$N ``saturation'' $S/A$-value of 7.9 is also shown in this figure. As can be seen, 
the total entropy increases steadily with collision energy.

\begin{figure}[htpb]
 \begin{minipage}[h]{0.5\linewidth}
    \includegraphics[width=70mm,angle=0]{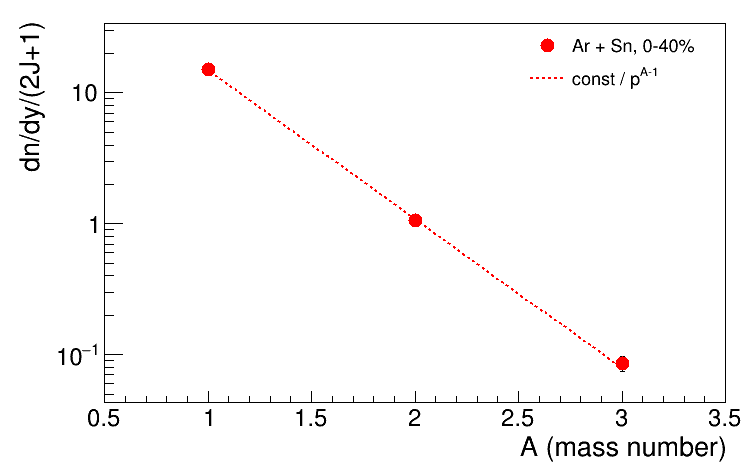}
 \end{minipage}
 \begin{minipage}[h]{0.5\linewidth}
   \includegraphics[width=70mm,angle=0]{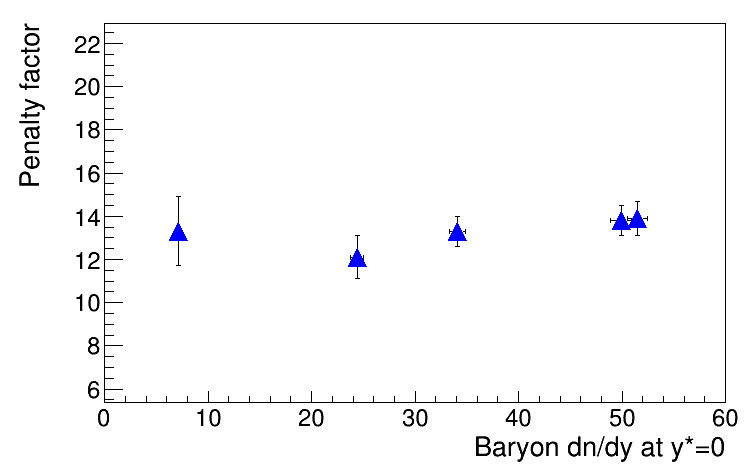}
 \end{minipage}
 \caption{Left: Midrapidity d$n$/d$y$/(2J+1) for $p,d,t$ from central Ar+Sn collisions.
 The dashed line is a fit to eq.~(\ref{eq13}). Right: Penalty factor from central Ar+A
 collisions versus baryon rapidity density at midrapidity.}
 \label{fig_yields_A}
\end{figure}

It has been established experimentally that the cluster production yields scaled by the spin 
degeneracy factor (2J+1) decrease exponentially with the atomic mass number $A$~\cite{E864,NA49_dtHe3}.
 As an example, d$n$/d$y$/(2J+1) at midrapidity for $p,d,t$ as a function of $A$ from 0--40\% 
 central Ar+Sn collisions are preseneted in  figure~\ref{fig_yields_A} (left panel). 
 The particle rapidity density values are extracted from the fits of figure~\ref{mtyields}. 
 The $A$-dependence of the yields was fitted to a form:
\begin{equation}\tag{13}
\label{eq13}
    \frac{dn}{dy}(A) = const / p^{A-1},
\end{equation}
 where the parameter $p$ (`penalty factor') determines the penalty for adding one extra nucleon to the system.
 
The $p$-factors from central Ar+A collisions are shown in figure~\ref{fig_yields_A} (right panel)
as a function of the midrapidity baryon rapidity density. The errors are the statistical errors
obtained from the fit to eq.~(\ref{eq13}).

Recently, the STAR experiment reported measurements of the compound yield ratio $R_{ptd} = N_p N_t/N_d^2$ 
of protons ($N_p$) and tritons
($N_t$) to deuterons ($N_d$)~\cite{STAR_pdt}. Coalescence models predict~\cite{QCDprob}
that a non-monotonic behavior of the ratio as a function of the system size or collision energy 
is a signature of the neutron density fluctuations $\Delta n$: $R_{ptd} \approx g(1+\Delta n)$ with 
a color factor $g\simeq 0.29$. Following this argument,  $R_{ptd}$ is a promising observable to
search for the critical point and/or a first-order phase transition in heavy-ion collisions~\cite{LightNucl}. 
In coalescence models, the compound
yield ratio should increase as the size of the system  decreases. Indeed, this effect is observed by 
the STAR experiment~\cite{STAR_LightNucl}. 
\begin{table}[!hbp]
\vspace{-0.5cm}

  \caption {
$N_p N_t/N_d^2$ values evaluated from the mean $dN/dy$ values of protons, deuterons and tritons 
over the rapidity range $-0.18<y^*<0.22$ and $0.22<y^*<0.62$ in Ar+A interactions with centrality 0--40\%. 
The quoted errors are the quadratic sums of the statistical and systematic uncertainties.  }
\begin{footnotesize}
\vspace{0.3cm}
\begin{tabular}{|c|c|c|c|c|c|}
\hline
& & & & & \\
& Ar+C & Ar+Al & Ar+Cu & Ar+Sn & Ar+Pb \\
& & & & & \\
\hline
& & & & & \\
$N_p N_t/N_d^2$    & $0.52\pm0.18$  & $0.53\pm0.10$ & $0.66\pm0.16$ & $0.68\pm0.12$  & $0.57\pm0.11$  \\
($-0.18<y^*<0.22$) & & & & & \\
$N_p N_t/N_d^2$    &  --   & $0.40\pm0.07$ & $0.60\pm0.08$ & $0.50\pm0.08$  & $0.51\pm0.12$  \\
($0.22<y^*<0.62$)& & & & & \\
\hline
\end{tabular}
\end{footnotesize}
\label{table_NpNt_toNd2}
\end{table}
\vspace{0.5cm}

\begin{figure}[htbp]
\begin{center}
\vspace{-1.0cm}

\includegraphics[width=0.99\textwidth,bb=0 0 1382 653]{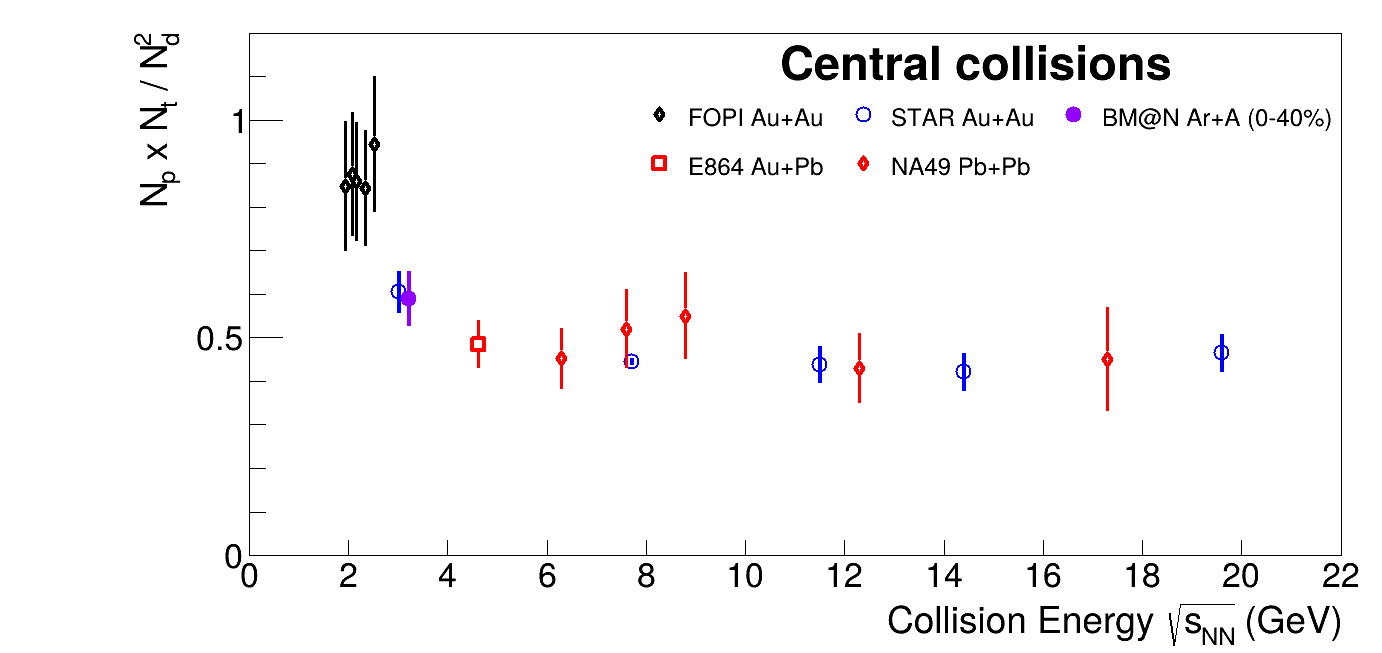}
 \end{center}
\vspace{-0.5cm}
\caption{Compound yield ratio $N_p\cdot N_t/N_d^2$ of protons ($N_p$) and tritons
($N_t$) to deuterons ($N_d^2$) as a function of the center-of-mass energy of nucleus-nucleus interactions.
The BM$@$N result represents the weighted average value in the rapidity range $-0.18<y^*<0.22$ calculated 
for Ar+Al, Cu, Sn and Pb interactions with centrality 0--40\%.}
\label{NpNtNd2_compare}
\end{figure}

To evaluate the $R_{ptd}$ ratio, mean values of the $dN/dy$ distributions for protons, deuterons and 
tritons are calculated in two rapidity ranges:  $0.9<y<1.3$ ($-0.18<y^*<0.22$) and $1.3<y<1.7$ ($0.22<y^*<0.62$). 
The results are given
in table~\ref{table_NpNt_toNd2} for argon-nucleus interactions with centrality 0--40\%. 
 
No significant variation of the $N_p N_t/N_d^2$
values is observed with the various targets. Taking the differences as systematic uncertainties, 
the weighted average value of the compound ratio is estimated to be $0.59\pm 0.065$ for $-0.18<y^*<0.22$ 
and $0.46\pm 0.10$ for $0.22<y^*<0.62$, 
 where the uncertainty is the quadratic sum of the statistical and systematic uncertainties. Within the 
 uncertainties, there is no strong dependence of the  $R_{ptd}$ ratio on rapidity in the measured rapidity 
 range. The BM$@$N value for $R_{ptd}$  for $-0.18<y^*<0.22$ is compared in figure~\ref{NpNtNd2_compare} 
 with the measurements of other experiments. The BM$@$N result lays between the values of 0.8--1.0 derived 
 by the FOPI experiment (impact parameter $b_0 < 0.15$)~\cite{FOPI_pdt} at lower energies and the values
 of 0.4--0.5 obtained by the E864 (0--10\% central) \cite{E864}, STAR (0--10\% central) 
 \cite{STAR_pdt,STAR_LightNucl} and NA49 (0--7\% central) \cite{NA49_dtHe3} experiments at higher CM energies
  from 4.3 to 18 GeV. The BM$@$N value for $R_{ptd}$ is consistent with the STAR Au+Au result measured in 
  the fixed target mode at $\sqrt{s_{NN}}$ of 3~GeV~\cite{STAR_PHQMD}.

\section{Conclusions}
\label{sect7}

The first results of the BM$@$N experiment are presented on the proton, deuteron and triton
yields and their ratios in argon-nucleus interactions at the beam kinetic energy of 3.2~A~GeV.
They are compared with the DCM-SMM and PHQMD models and with previously published results
of other experiments.

The transverse mass $m_T$ spectra are measured and the mean transverse kinetic energy 
$\langle E_T\rangle=\langle m_T\rangle -m$ are presented for more  central 0--40\% events as 
functions of the rapidity $y$ and  mass $m$ of the nuclear fragment.  The $\langle E_T\rangle$ values
are found to depend linearly on the mass $m$. The source temperature at kinetic freeze-out and 
the average radial velocity are extracted  within the Blast-Wave model.

The rapidity density $dN/dy$ of protons, deuterons and tritons are presented for the whole $p_T$ 
range in two centrality ranges.  The DCM-SMM and PHQMD  models reproduce the shapes of the
spectra but underestimate the deuteron yields by  factors of about 3 and 2, respectively.  
The triton yields predicted by the models are below the experimental data by a factor of about 5.

The average rapidity loss $\langle \delta y \rangle$ increases with the target mass and with the 
collision centrality.  In contrast, the rapidity loss scaled to the beam rapidity 
$\langle \delta y \rangle$/$y_b$ in almost symmetric heavy-ion collisions does not vary significantly 
over a broad energy range.

The ratio of deuterons to protons $R_{dp}$ rises in peripheral collisions and levels off in
central ones, possibly indicating a saturation of the nucleon phase-space density at 
freeze-out. The entropy per baryon $S/A$ was estimated to be $S/A\approx 8$ nicely fitting in 
the trend of the $S/A$ energy dependence established from other experimental results.  

The proton, deuteron and triton yields are used to calculate the coalescence parameters $B_2$ and $B_3$ 
for deuterons and tritons. 
Consistent coalescence radii are extracted from $B_2$ and $B_3$ values extrapolated to $p_T=0$. 
They are slightly lower compared with the results of experiments at higher energies in agreement with 
a weak increase of the coalescence radii with increasing collision energy.
 
The compound yield ratio $N_p N_t/N_d^2$ of protons and tritons to deuterons is evaluated and compared 
with other measurements at lower and higher energies.
The results follow the general trend of decreasing values of $B_2$, $B_3$ and $N_p N_t/N_d^2$ ratio with increasing energy.

\paragraph{Acknowledgments.}
The BM$@$N Collaboration acknowledges the efforts of the staff of the accelerator division of the
Laboratory of High Energy Physics at JINR that made this experiment possible.
The BM$@$N Collaboration acknowledges support of the HybriLIT of JINR for the provided computational 
resources. The research has been supported by the Ministry of Science and Higher Education of the 
Russian Federation, Project ``New Phenomena in Particle Physics and the Early Universe'' 
No. FSWU-2023-0073 and by the Science Committee of the Ministry of Science and Higher Education of the 
Republic of Kazakhstan (Grant No. AP23487706).

%
%

\end{document}